\documentclass[fleqn,usenatbib]{mnras}
\pdfoutput=1

\usepackage{newtxtext,newtxmath}

\usepackage[T1]{fontenc}
\usepackage{ulem}

\DeclareRobustCommand{\VAN}[3]{#2}
\let\VANthebibliography\thebibliography
\def\thebibliography{\DeclareRobustCommand{\VAN}[3]{##3}\VANthebibliography}

\usepackage{multicol}

\usepackage{graphicx}	
\usepackage{amsmath}

\newcommand{\msol}{M$_\odot$}

\title[Constraining nucleonic dense matter parameters]{Constraining a relativistic mean field model using neutron star mass-radius measurements I: Nucleonic models
}

\author[Huang et al.]{
Chun Huang,$^{1,2,3}$\thanks{E-mail: chun.h@wustl.edu}
Geert Raaijmakers,$^{4}$
Anna L. Watts,$^{2}$
Laura Tolos,$^{5,6,7}$
and Constan\c{c}a Provid\^{e}ncia,$^{8}$
\\
$^{1}$Physics Department, Central China Normal University, Luoyu Road, 430030, Wuhan, China \\
$^{2}$Anton Pannekoek Institute for Astronomy, University of Amsterdam, Science Park 904, 1090 GE Amsterdam, the Netherlands\\
$^{3}$Physics Department, Washington University, One Brookings Drive, St. Louis, MO, 63130, USA \\
$^{4}$GRAPPA, University of Amsterdam, Science Park 904, 1098 XH Amsterdam, The Netherlands \\
$^{5}$Institute of Space Sciences (ICE, CSIC), Campus UAB, Carrer de Can Magrans, 08193, Barcelona, Spain \\
$^{6}$Institut d'Estudis Espacials de Catalunya (IEEC), 08034, Barcelona, Spain \\
${^7}$Frankfurt Institute for Advanced Studies, Ruth-Moufang-Str. 1, 60438, Frankfurt am Main, Germany \\
$^{8}$CFisUC, Department of Physics, University of Coimbra, 3004-516 Coimbra, Portugal
}

\date{Accepted XXX. Received YYY; in original form ZZZ}

\pubyear{2023}

\begin{document}
\label{firstpage}
\pagerange{\pageref{firstpage}--\pageref{lastpage}}
\maketitle

\begin{abstract}
Measurements of neutron star mass and radius or tidal deformability deliver unique insight into the equation of state (EOS) of cold dense matter.  EOS inference is very often done using generalized parametric or non-parametric models which deliver no information on composition.  In this paper we consider a microscopic nuclear EOS model based on a field theoretical approach.  We show that current measurements from NICER and gravitational wave observations constrain primarily the symmetric nuclear matter EOS.  We then explore what could be delivered by measurements of mass and radius at the level anticipated for future large-area X-ray timing telescopes. These should be able to place very strong limits on the symmetric nuclear matter EOS, in addition to constraining the nuclear symmetry energy that determines the proton fraction inside the neutron star.
\end{abstract}


\section{Introduction}

Neutron star observations, via electromagnetic radiation or gravitational waves (GWs), provide unique insight into the dense matter Equation of State (EOS), particularly at low temperatures \citep{Lattimer12ARNPS,OE17,Baym2018,Tolos2020,YangPiek2020,Hebeler2021,Burgio:2021vgk}.  Measurements of the highest neutron star masses already restrict the range of dense matter possibilities \citep[see e.g.][]{Ozel2010,Cromartie2020}. GW detectors are now providing constraints on tidal deformability from binary neutron star mergers \citep{2018PhRvL.121p1101A,GW170817_TD2,GW190425}, and NASA's Neutron Star Interior Composition Explorer \cite[NICER,][]{Gendreau2016} has reported its first results for the simultaneous inference of mass and radius \citep{Miller2019,Riley2019,Miller2021,Riley2021,Salmi22,Vinciguerra24}.  These multi-messenger results are now being combined, together with information from laboratory nuclear experiments, to constrain EOS models \citep[for a selection of recent papers see][]{Miller2021,Raaijmakers_2021,JieLiJ21,Legred21,Pang21,TangSP21,Annala2022,Biswas2022}.

Given our lack of knowledge of the EOS relating to high densities and very asymmetric matter, or the appearance of exotic degrees of freedom, the community frequently uses meta-models that cover the whole acceptable pressure-energy density or mass-radius (M-R) domains. Different methodologies have been used to implement this strategy, for example through the parametrization of the EOS via polytropes \citep{Read2009,Kurkela:2014vha,Most2018,Annala18}, the speed of sound \citep{Alford:2013aca,Bedaque2015,Tews2018,Greif_2019,Annala2020} or spectral representations \citep{Lindblom2010,2018PhRvL.121p1101A, Lindblom2022} imposing, if necessary, causality. More recently, a non-parametric approach was also proposed through the introduction of Gaussian processes \citep{Landry:2020vaw,Essick2021,Legred2022}. 

While these approaches allow us to determine the EOS domain that satisfies the observational constraints, they deliver no information on composition, either the proton fraction or the existence of non-nucleonic degrees of freedom such as hyperons, delta-baryons or deconfined quark matter. In the present study, the drawbacks of meta-models will be overcome by considering a microscopic model based on a relativistic field theoretical approach to determine the whole EOS space allowed by observations. This will be undertaken by changing the model parameters. Being defined in a relativistic framework, the model automatically incorporates causality.  This approach has already been taken in \citet{Traversi:2020aaa}, where the relativistic mean-field model (RMF) of \citet{BOGUTA1977413} was applied; and in \citet{Malik:2022zol} who considered as their underlying model an RMF with density dependent couplings. The authors of \citet{Sun2022} have recently developed a Bayesian inference approach, in the framework of several nuclear RMF, to determine how GW and NICER measurements constrain the $\Lambda-\sigma$ and  $\Lambda-\omega$ couplings, while fixing the $\Sigma$ and $\Xi$ couplings to reasonable values. A major advantage of this methodology is the possibility, once the inference is completed, to discuss the  composition of matter, in particular, the proton fraction,  or the nuclear matter properties.  In the present study, we will base our approach on the framework that is behind the reasonably successful FSU2R nucleonic model \citep{Tolos_2016,Tolos_2017}.

The power of astrophysical observations to constrain microphysical models is also important to inform the design and observational strategy for future X-ray telescopes that will exploit the M-R inference technique used by NICER \citep{Watts_2016,Watts2019}.  Large-area X-ray spectral timing telescope concepts like the {\it enhanced X-ray Timing and Polarimetry mission} \citep[eXTP, ][]{extp_zhang,extp_watts} and the {\it Spectroscopic Time-Resolving Observatory for Broadband Energy X-rays} \citep[STROBE-X,][]{strobex} aim to measure mass and radius for not only faint rotation-powered millisecond pulsars (the sources being studied by NICER) but also accreting neutron stars.  These sources are often transient, as are many of the other high priority targets (such as black hole binaries) for these missions, and long observation times are required to build up sufficient photons to enable the analysis.  Being able to make a well-informed decision on the potential scientific pay-off of observing a particular source is important, particularly if it comes at the expense of observing another potentially attractive target.  

In this paper we therefore consider not only the constraints arising from existing observations, but look at what might be delivered by future missions.  Since we are particularly interested in determining the model space constrained by the astrophysical observations, we therefore impose only a minimal number of nuclear matter properties. A different approach was undertaken in \cite{Ghosh_2022_1,Ghosh_2022} where, within a cut off scheme applied to a prior already constrained by nuclear matter properties, constraints from chiral Effective Field Theory, Heavy Ion Collisions and astrophysics were imposed as filters.

In Section \ref{EOS} we introduce the EOS model and our choice of priors.  In Section \ref{inference} we describe the Bayesian inference procedure and the M-R scenarios that we consider in our analysis.  Section \ref{results_now} gives the results of inference using currently available M-R and tidal deformability constraints, while Section \ref{results_future} considers what could be achieved by observations of M-R with future large-area X-ray telescopes.  In Section \ref{physics} we discuss the implications of our findings,  while we present our conclusions in Section \ref{sec:conclusions}.

\section{Equation of State models}
\label{EOS}

The EOS model we use for inference is the RMF of matter, where nucleons interact through the exchange of mesons and which provides a covariant description of the EOS and nuclear systems, using the parametrization denoted as FSU2R for nucleonic matter \citep{Tolos_2016,Tolos_2017}.  This model was developed from the nucleonic FSU2 model of \citet{Chen_2014} which was optimized to describe a set of properties of finite nuclei and of neutron stars. In particular, FSU2R has been developed to describe two solar mass stars and stellar radii below 13 km. 

The  well calibrated parameter set of the FSU2R model is chosen as the central value of our prior distributions (see Section \ref{subsec:EOS-prior}). There are presently many parametrizations based on the same framework including Z272v \citep{Horowitz:2000xj,Pais:2016xiu}, FSU \citep{Todd-Rutel:2005yzo}, IUFSU \citep{Fattoyev:2010mx,Cavagnoli:2011ft}, TM1$\omega\rho$ \citep{Providencia:2012rx,Bao:2014lqa}, TM1e \citep{Shen:2020sec}, TM1-2$\omega\rho$ \citep{Providencia:2012rx} and Big Apple \citep{Fattoyev:2020cws}.

The different models are based on distinct specific subsets of parameters chosen  by different calibration methods. Thus, even though our inference is based on FSU2R, the power of the present inference approach extends far beyond this scheme, and could provide constraining and excluding evidence for numerous models that are based on the RMF description for nucleonic matter.
			
In the following subsection, we formulate the RMF framework \citep{Serot:1984ey,Glendenning:1997wn}, then introduce our choice of priors for the full set of EOS parameters, which reproduce currently known nuclear physics quantities. For this last step we consider the relations between the EOS parameters and nuclear matter properties \citep{Chen_2014}.   In the last subsection we calculate the M-R prior based on our set-up. We then apply the model to the inference and Bayesian constraint process in Section \ref{inference}.
			
\subsection{Equation of State}
\label{eos}
The starting point of our theoretical framework is the Lagrangian density, which will be divided into three parts: nucleonic Lagrangian density $N$, lepton contribution $l$ (e,$\mu$) and meson field terms $\mathcal{M}$ ($\sigma$, $\omega$ $\rho$):
\begin{equation}
\mathcal{L}=\sum_{N}\mathcal{L}_{N} + \mathcal{L}_{\mathcal{M}} +\sum_{l}\mathcal{L}_{l}.
\end{equation}
They can be separately expressed as
\begin{equation}
\begin{aligned}
\sum_{N}\mathcal{L}_{N} &= \sum_{N}
\bar{\Psi}_{N}\left( i\gamma_{\mu}\partial^{\mu} 
- m_{N} + g_{\sigma }\sigma \right.\\
&\left.  - g_{\omega }\gamma_{\mu}\omega^{\mu}
- g_{\rho }\gamma_{\mu}\vec{I}_{N}\cdot \vec{\rho}^{\mu}
\right)\Psi_{N} ,
\\
\sum_{l}\mathcal{L}_{l}&= \sum_{l}\bar{\psi}_{l}( i\gamma_{\mu}\partial^{\mu} 
-m_{l})\psi_{l} , \\
\mathcal{L}_{\mathcal{M}}&=\frac{1}{2}\partial_{\mu}\sigma\partial^{\mu}\sigma-\frac{1}{2}m_{\sigma}^{2}\sigma^{2} -\frac{\kappa}{3!}(g_{\sigma } \sigma)^{3}-\frac{\lambda_{0}}{4!}(g_{\sigma } \sigma)^{4}\\
&-\frac{1}{4}\Omega^{\mu\nu}\Omega_{\mu\nu}
+\frac{1}{2} m_{\omega}^2 \omega_{\mu}\omega^{\mu}+\frac{\zeta}{4!} g_{\omega }^4 (\omega_{\mu}\omega^{\mu})^{2} \\
&-\frac{1}{4}\vec{R}^{\mu\nu}\cdot\vec{R}_{\mu\nu}+\frac{1}{2}m_{\rho}^{2}\vec{\rho}_{\mu}\cdot\vec{\rho}^{\mu}+\Lambda_{\omega}g_{\rho }^{2}\vec{\rho}_{\mu}\cdot\vec{\rho}^{\mu}g_{\omega }^{2}\omega_{\mu}\omega^{\mu} ,
\end{aligned}
\end{equation}
where $\Psi_{N}$ and $\psi_{l}$ are the nucleon  and lepton spinors, and $\vec{I}_{N}$ is the nucleon isospin operator. The coupling of a meson to a nucleon is denoted by $g$, while the masses of the nucleons, mesons, and leptons are denoted by $m$. The parameters $\kappa$, $\lambda_{0}$, $\zeta$ and $\Lambda_{\omega}$ plus the meson-nucleon coupling constants are coupling constants to be determined by the inference method.

By solving the Euler-Lagrange equations for nucleons and mesons in the RMF approach, we can obtain the Dirac equation for nucleons and the equations of motion for the expectation values of the mesons, as explained in \citet{Dutra:2014qga}. The EOS and composition of matter are then obtained by coupling those equations with global charge neutrality and $\beta$-equilibrium conditions,  which relates the chemical potentials of the different species.

\subsection{Choice of priors for model parameters}
\label{subsec:EOS-prior}
The free parameters in the EOS contain unique information on the field coupling strengths, meson mass or interaction. The word `free' indicates that their values have to be determined, and could vary within a reasonable range given by ground-based experiments. The parameters in the nucleonic EOS can be divided into three groups:
\begin{enumerate}

\item scalar self-interaction coupling constants and mixed-interaction constants among mesons:
\begin{equation}
\kappa\qquad \lambda_{0} \qquad \zeta \qquad \Lambda_{\omega} .
\end{equation}
The parameters $\kappa$ and $\lambda_{0}$ are introduced to indicate the $\sigma$ meson self-interaction \citep{BOGUTA1977413}; these two are constrained by reproducing the equilibrium properties of symmetric nuclear matter and finite nuclei. Both of them are responsible for producing the incompressibility $K$ \citep{BOGUTA1977413,BOGUTA1983289,Mueller:1996pm} in agreement with 
experiments on nuclear giant resonances and heavy ion collisions. The parameter $\zeta$ is the quartic self-coupling of the $\omega$ meson \citep{BODMER1991703}, and heavily influences the high density behavior of the EOS dominating the largest mass stars, as clearly discussed in \citet{Mueller:1996pm} already in the nineties.  Moreover, there is a mixed-interaction term, $\Lambda_{\omega}$ \citep{Horowitz:2000xj}, between the $\omega$ and $\rho$ meson. This term is responsible for modifying the density dependence of the nuclear symmetry energy, and influences the neutron-skin radius of heavy nuclei and the radii of neutron stars first discussed in \citet{Horowitz:2000xj,Horowitz:2001ya} where it was shown that the larger $\Lambda_{\omega}$, the smaller the radius of the canonical neutron star and the  neutron radius of a heavy nucleus. The joint effect of both parameters on the neutron star properties is nicely illustrated in \citet{Fattoyev:2010rx}: $\zeta$ controls 
the maximum neutron star mass and $\Lambda_{\omega}$ has an effect on the radius of intermediate and high mass stars\footnote{In the following we will refer to low, intermediate and  high mass stars, as stars that have respectively a mass below $1.4\, M_\odot$, a mass in the range  $1.4\lesssim M/M_\odot\lesssim 2$, a mass $\gtrsim 2M_\odot$.} (for a similar discussion see \citealt{Fattoyev:2010tb}). This last parameter defines the slope of the symmetry energy at saturation, $L$, as shown in \citet{Cavagnoli:2011ft}, where a correlation between $L$ and the radius of 1.0 and 1.4$M_\odot$ stars was verified. However, in \citet{Alam:2016cli} the authors show that this correlation is strong only for low mass stars and becomes weaker as the mass increases. For masses above 1.4$M_\odot$ the correlation of the radius with the nuclear matter incompressibility is stronger. A similar conclusion concerning the correlation between the slope $L$ and the neutron star radius was drawn in \citet{Fortin:2016hny} and \citet{Malik:2018zcf}.

\item meson-nucleon coupling constants:
\begin{equation}
g_{\sigma }\qquad g_{\omega } \qquad g_{\rho } .
\end{equation}
The parameters $g_{\sigma }, g_{\omega }$ are the couplings between the nucleon and the isoscalar $\sigma$ and $\omega$ mesons, respectively. Those determine the energy per particle and density of the nuclear matter saturation point, thus becoming instrumental for the ground-state properties of finite nuclei. The $g_{\rho }$ represents the coupling constant of the isovector $\rho$ with the nucleon, which is responsible for producing a reasonable nuclear symmetry energy, impacting the properties of heavy neutron-rich nuclei and of neutron stars.
\item meson mass
\begin{equation}
m_{\sigma}\qquad m_{\omega}\qquad m_{\rho} .
\end{equation}
These are the masses of the $\sigma$, $\omega$ and $\rho$ meson, respectively. The values of $m_{\omega}$ and $m_{\rho}$ are well determined (782.5 MeV and 763 MeV, respectively), but $m_{\sigma}$ is less well-established, with a range from 495 MeV to 510 MeV. 

As discussed in \cite{Glendenning:1997wn}, infinite nuclear matter described by the present model depends on the coupling constants and the meson masses only through the ratios $g_i/m_i$. We have confirmed that our results do not depend on the $\sigma$-meson mass and have set it to a constant with a value of 497.479 MeV, consistent with the FSU2R  model.
\end{enumerate}
			
Taking into account all parameters, the nucleonic model space is a seven dimensional parameter space. However, each parameter can have its own distribution. Using some reasonable values of the parameter space based on nuclear experimental constraints, we make the following choices for the prior distributions. 

\begin{table}
\centering
\setlength{\tabcolsep}{11mm}{\begin{tabular}{l c} 
\hline\hline
\text {EOS parameter} & {Prior} \\
\hline
$\kappa$ (MeV)&$N(2.525, 1.525^2)$\\ $\lambda_{0}$ & $N(0.0045,0.0205^2)$\\
$\zeta$ & $\mathcal{U}(0,0.04)$\\
$\Lambda_{\omega}$&$\mathcal{U}(0,0.045)$\\ 
$g_{\sigma  }^{2}$&$N(107.5, 7.5^{2})$\\
$g_{\omega }^{2}$& $\mathcal{U}(150, 210)$\\ $g_{\rho }^{2}$&$ \mathcal{U}(75,210)$\\
[1ex] 
\hline\hline
\end{tabular}}
\caption{This is a summary for all the EOS parameters prior setting, where $N$ stands for Gaussian distribution and $\mathcal{U}$ means Uniform (Flat) distribution.}
\label{table:1}
\end{table}

We start with the parameter range for the scalar self-interaction coupling constants and mixed-interaction constants. The parameter $\kappa$ is set as a Gaussian distribution, centered at 2.525 with $\sigma = 1.525$. The 1-$\sigma$ credible interval thus covers the range 1 to 4.05, denoted as $N(2.525, 1.525^{2})$ ($N$ for Gaussian distribution). 
The parameter $\lambda_{0}$ also has a Gaussian prior with -0.016 $\sim$ 0.025 as the $\pm$ 1-$\sigma$ range, centered at 0.0045, denoted as $N(0.0045, 0.0205^{2})$. The $\zeta$ parameter must be non-negative to prevent abnormal solutions of the vector field equation of motion. The parameter space where $\zeta > 0.04$ falls outside of the scope of our investigation, because it does not permit neutron stars with masses that reach 2\msol, so we set a flat prior 0 $\sim$ 0.04, denoted as $\mathcal{U}(0,0.04)$ ($\mathcal{U}$ for flat distribution, 0 and 0.04 are the lower and upper limit). 
The final $\Lambda_{\omega}$ parameter should be also non-negative, and we use a flat prior from 0 to 0.045, $\mathcal{U}(0,0.045)$; outside this range the prior probability is set to zero.
			
As for the meson-nucleon couplings, the favored value ranges are sometimes different depending on the experiment \citep{Dutra:2014qga}. Due to this, wide ranges for these quantities are chosen and a hard cut-off is never used for the distributions of the parameters. The $g_{\sigma  }^{2}$ prior is a Gaussian distribution $N(107.5, 7.5^{2})$. For $g_{\omega }^{2}$ we choose a flat prior, from 150 to 210, denoted as $\mathcal{U}(150,210)$. The $g_{\rho }^{2}$ distribution is set as flat prior $\mathcal{U}(75,210)$.
			
Together, these define a seven-dimensional prior space for the EOS (see Table \ref{table:1}), from which we can sample.
			
\subsection{Nuclear matter saturation properties}
\label{nucsatprop}
\begin{figure*}
\centering
\includegraphics[scale=0.4]{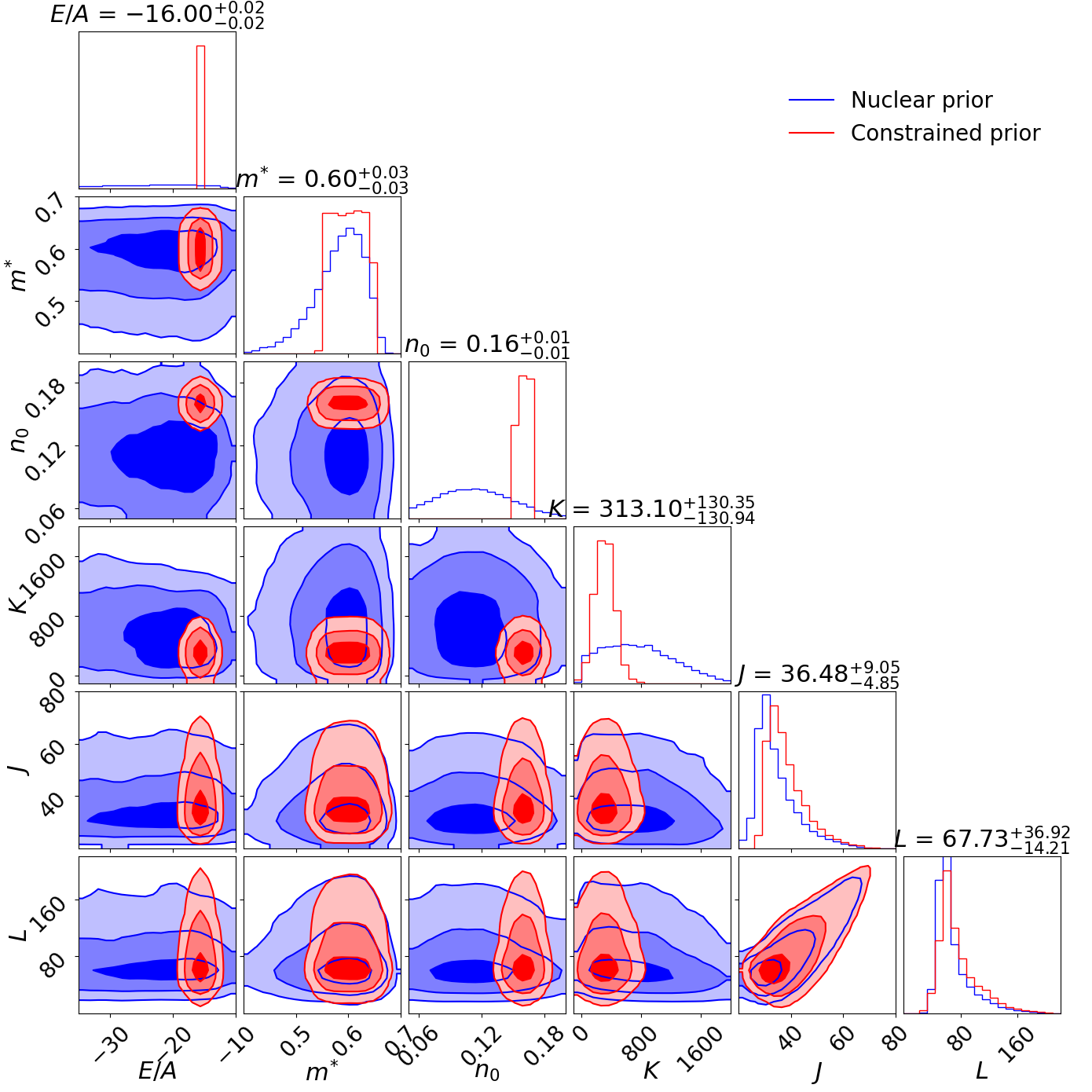}
\caption{This corner plot illustrates the posterior distributions of the nuclear saturation properties after introducing constraints on $E/A$, $m^*$, $n_0$ and the incompressibility of symmetric nuclear matter $K$ (see Section \ref{nucsatprop}) compared to the prior space on these properties established by the priors on the EOS parameters.  $K$ in particular would otherwise have a wide prior range that is much broader than all experimentally allowed values (note that formally the full EOS parameter priors in Table \ref{table:1} also admit values of $K$ that are negative, but these are unphysical so we do not show them in this plot). $J$ is the symmetry energy at saturation density, and $L$ is the slope of the symmetry energy at saturation density. The corner plot was produced by randomly sampling the EOS parameter space using 100,000 samples, and then using the relationships between the EOS parameters and various nuclear matter properties to project the sampled points onto a five-dimensional space of nuclear quantities. In other words, the corner plot represents the distribution of nuclear quantities obtained by sampling the EOS parameter space and using the relevant relationships to compute the corresponding nuclear matter properties. The contour levels in the corner plot, going from deep to light colors, correspond to the 68\%, 84\%, and 98.9\% levels. }
\label{nuclear}
\end{figure*}

All the parameters from the RMF model have a direct connection with the nuclear saturation properties. The relation between the parameters and the nuclear matter saturation properties can be found, for example, in the Appendix of \citet{Chen_2014}.

As shown in that paper, given a set of isoscalar and isovector parameters, we can compute the nuclear saturation density ($n_0$), the energy per nucleon of symmetric matter at saturation density ($E/A$), the nucleon effective mass ($M^*$) and  incompressibility of symmetric nuclear matter at saturation density ($K$), as well as the symmetry energy at saturation density ($J$) and its slope at that density ($L$). These nuclear matter saturation properties are reasonably well-constrained by experiments, so that a plausible range is known for some of them \citep{Dutra:2014qga,OE17,Margueron:2017eqc,Huth:2020ozf}. In particular, the nuclear saturation density  $n_0$ ranges from 0.15 to 0.17 fm$^{-3}$, the binding energy per particle $E/A$ from 15.8 to 16.2 MeV, the incompressibility $K$ from 175 to 315 MeV \citep{Huth:2020ozf,Stone:2014wza}, the effective mass $M^{*}/m$ goes from 0.55 to 0.65 \citep{Hornick:2018kfi}, and the range for the symmetry energy $J$ lies between 25 and 38 MeV, whereas for the slope of the symmetry energy $L$ we have values between 30 and 86 MeV \citep{Lattimer:2012xj,OE17}.  

Note that a larger slope has been determined from the measurement of the skin thickness of $^{208}$Pb \citep{PREX:2021umo}, in particular, $L=106 \pm$ 37 MeV was estimated in \citet{Reed2021}. However, other studies \citep{Yue:2021yfx,Essick2021,Reinhard:2021utv} have obtained smaller values of $L$. As has been discussed in \citet{Reinhard:2021utv,Reinhard:2022inh,Mondal:2022cva} there is currently still some tension between experiment and theory, and it is not clear whether the existing experimental uncertainties in the measured parity-violating asymmetry in PREX-II \citep{PREX:2021umo} make it an adequate observable to constrain theoretical models that have otherwise successfully described many other nuclear matter properties.
   
The values for the nuclear matter saturation properties that result from our choice of the parameter priors are spread over wider ranges than the values from nuclear experiments, especially for the incompressibility of symmetric nuclear matter $K$. We therefore impose some additional conditions: firstly, we fix $E/A$ to be a Gaussian distribution centered on -16 MeV, with $\sigma = 0.02$, $N(-16, 0.02^{2})$, and restrict $m^{*}$ as $\mathcal{U}(0.55,0.64)$ and $n_0$ as $\mathcal{U}(0.15,0.17)$ as shown in Fig.\ref{nuclear}.  These quantities are needed to compute to $K$, $J$ and $L$. In order to include some minimal guidance for the parameter choices from the known nuclear physics properties, but still keeping the initial objective of constraining the EOS primarily through astronomical observations, we also include a loose prior condition on $K$. We impose the condition that the incompressibility should satisfy $100\lesssim K \lesssim 400$ MeV. To improve the convergence speed of the inference, we define a probability function as follows: $p(K) = -0.5\times|250 - K|^{10}/150^{10}$, which is a super-Gaussian function.  This is less extreme than a hard cut, but strongly disfavours values outside the nominal range. This condition still leaves enough freedom to explore the power of the astrophysical constraints, whilst avoiding extremely unreasonable nuclear matter properties.
			
The posterior distribution of nuclear quantities compared to the nuclear quantity prior generated from our defined EOS parameter prior is illustrated in Figure \ref{nuclear}. By applying these restrictions and utilizing the posterior of the EOS model parameters as new priors (referred to hereafter as {\it constrained priors}), it is possible to evaluate the constraining power of our observations while still maintaining the credibility of our inference from a nuclear physics perspective. 

The restriction on $K$ in particular has multiple effects on the different EOS model parameters due to the non-linear relationship between these parameters and nuclear matter properties. Some of the EOS parameter space of $\lambda_0$ has been excluded, as $\lambda_0$ exerts a significant influence on the radius of a 1.4 \msol~star, and is therefore of interest in the context of our investigation. The cut-off also has an impact on the parameter $\zeta$, shifting it towards larger values. This is interesting as larger values of $\zeta$ may produce maximum mass stars below 2 \msol, while smaller values are preferred for producing maximum mass stars above 2 \msol.  These opposing effects on $\zeta$ can lead to a strong constraint on the EOS parameters. This highlights the value of constraining these parameters using both nuclear physics and astrophysical methods. A similar shift can also be seen in the distributions of $g_{\omega}$ and $g_{\rho}$. The constrained priors (to be compared to the original priors in Table \ref{table:1}) can be seen in Figure \ref{J0740}.
   
\subsection{Mass-radius priors}
\label{subsec:M-R-prior}
\begin{figure}
\includegraphics[width=\linewidth]{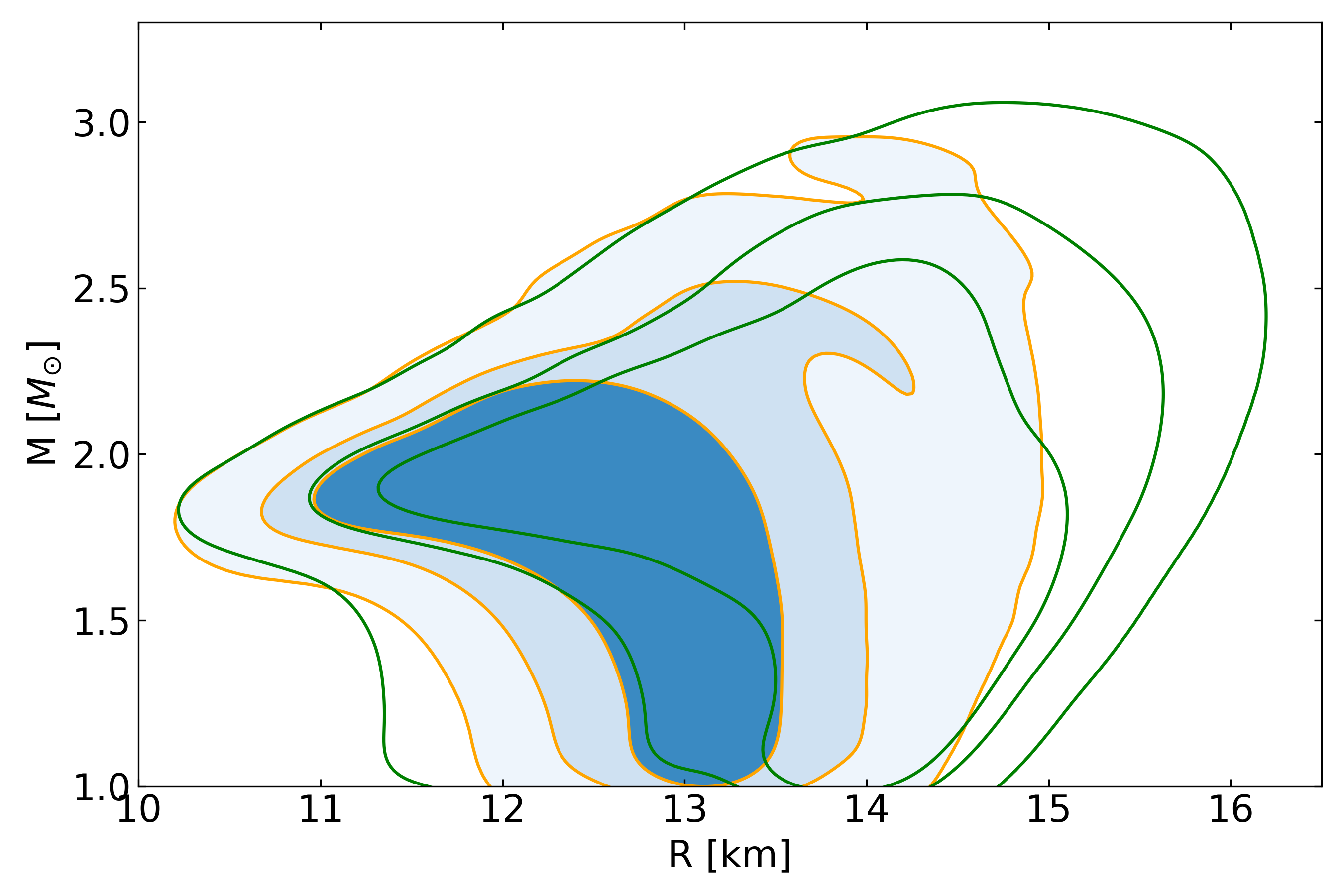}
\caption{The M-R posterior after applying all four nuclear saturation property constraints (orange/blue), compared to the M-R prior resulting from the initial EOS model with the full EOS priors but after imposing priors on $E/A$, $m^{*}$ and $n_0$ (green), to demonstrate the effect of the additional constraint on $K$. The contour levels, from the innermost to the outermost, correspond to the 68\%, 84\% and 100\% credible regions, 100\% being the point beyond which there are no more samples (for both the orange/blue and the green contours). After applying the $K$ constraint, there are no samples with radii above 16 km. Note that the prominences on the high radius side in the post-$K$ constraint contours are sampling artefacts.}
\label{M-R_nu} 
\end{figure}

Every point in EOS parameter space should be uniquely correlated to an EOS curve in the P-$\varepsilon$ plane. Then by varying central density, EOS points can be mapped to the M-R plane.  Understanding the M-R prior that results from the EOS model choices is vital when trying to infer the M-R relation from e.g. astrophysical measurements of M-R \citep{Greif_2019}. 

The M-R relation is derived by solving the Tolman-Oppenheimer-Volkoff (TOV) equations \citep{Oppenheimer39,Tolman39}. The TOV equations for a  static and spherically-symmetric system are 
\begin{equation}
\begin{aligned}
\frac{d P}{d r} &=-\frac{G}{r^{2}}(\varepsilon+P)\left(m+4 \pi r^{3} P\right)\left(1-\frac{2 G m}{r}\right)^{-1}, \\
\frac{d m}{d r} &=4 \pi r^{2} \varepsilon ,
\end{aligned}
\end{equation}
where $m$ is the star mass for a given radial coordinate $r$ (in spherical coordinates) and $G$ is the gravitational constant. The TOV equations are solved by spanning the possible central densities of a neutron star. For a given central density, this set of equations gives a unique solution, that is, a point in the M-R diagram. Repeating this process for different central densities, we can map out the M-R relation. This extends to a maximum mass, beyond which solutions to the TOV equation are unstable, at which point the relation is truncated.
			
To solve the TOV equations, we need to cover a wide range of densities. The RMF EOS introduced in Section \ref{eos} is taken to be the EOS in the neutron star core.  We need to match this to an EOS for the neutron star crust and carefully treat the interface region. We use the BPS EOS \citep{Baym:1971pw} for the outer crust, the region where the energy density runs from $\varepsilon_{\text{min}} = 1.0317 \times 10^{4} \mathrm{~g} / \mathrm{cm}^{3}$ to $\varepsilon_{\text {outer }} = 4.3 \times 10^{11} \mathrm{~g} / \mathrm{cm}^{3}$.  In the region between the core and the outer crust, $\varepsilon_{\text {outer }}<\varepsilon<\varepsilon_{c}$, where $\varepsilon_c$ defines the crust-core transition, we use a polytropic EOS to avoid the complexity of the pasta structure \citep{Carriere:2002bx,Piekarewicz:2014lba}.
 The complete EOS is then given by
\begin{equation}
P(\varepsilon)= \begin{cases}P_{\mathrm{BPS}}(\varepsilon), & \text { for } \varepsilon_{\text {min}}\leq \varepsilon \leq \varepsilon_{\text {outer }} \\ A+B \varepsilon^{4 / 3}, & \text { for } \varepsilon_{\text {outer }}<\varepsilon \leq \varepsilon_{c} \\ P_{\mathrm{RMF}}(\varepsilon), & \text { for } \varepsilon_{c}<\varepsilon\end{cases}
\label{inner}
\end{equation}
where $P_{\mathrm{RMF}}$ indicates the EOS computed from the RMF framework introduced previously. The parameters $A$ and $B$ are chosen to ensure matching at $\varepsilon_{\text {outer }}$ and $\varepsilon_{c}$.
To improve the fitting accuracy, and having as reference the FSU2R EOS, the inner crust is built in the following way: Eq. (\ref{inner}) is considered  together with four points, one at the transition from the outer to the inner crust, $\varepsilon_\mathrm{outer}$, and three other points from the unified FSU2R inner crust EOS obtained in \cite{Providencia:2018ywl}. The parameters  $A$ and $B$ are chosen so that the inner crust EOS best fits the four points. The inner crust EOS is then matched to the core EOS at $\varepsilon_c\sim2.14\times 10^{14}\mbox{g/cm}^3$ such that the pressure is an increasing function of the energy density.
			
 The M-R prior can then be sampled by sampling the EOS parameter space and then solving the TOV equations for those parameters. 
 The green contour lines in Figure \ref{M-R_nu} show the  M-R prior that results from the nucleonic seven-parameter priors defined in Table \ref{table:1} together with the additional constraints on $E/A$, $m^*$ and $n_0$.  Our EOS model and prior choices favour radii in the range 13.5-14.5 km for a 1.4 \msol~neutron star, and 11.5 to 14.5 km for a 2 \msol~neutron star.   Note that our choices do not admit any solutions for radii below 10 km in the prior space. They do however admit some solutions above 16 km (the maximum permissible radius is 16.18 km), which is slightly larger than the maximum of 16 km assumed in the NICER Pulse Profile Modelling analysis of \citet{Riley2019, Riley2021,Salmi22}. 

After applying the $K$ constraint, however, the maximum radius is under 15 km (see Fig.~\ref{M-R_nu}) so that the M-R space is now consistent with the priors assumed in these NICER analyses. The maximum mass is also smaller than in the case where the $K$ constraint is not taken into account. This outcome is in line with expectations, since the $K$ constraint generally favors larger values of $\zeta$ and smaller values of $g_{\omega}$. This results in a generally softer EOS, that favors a smaller maximum mass and a smaller maximum radius.

\section{Inference framework}
  \label{inference}
We consider two types of inference scenario in this work. Firstly, we study the constraining power of current astronomical measurements: maximum masses derived from radio pulsar timing, GW measurements of tidal deformability, and M-R measurements from NICER. Secondly, we investigate the constraining power anticipated for M-R measurements made by future X-ray telescopes, like STROBE-X and eXTP. We will refer to these two scenarios as the { \it current constraints} and {\it future constraints} scenarios, respectively. Our goal with these analyses is to clarify both the constraining power we have now for fundamental nucleonic model parameters, and the prospects offered by more powerful X-ray telescopes.
           
The Bayesian inference methodology for the EOS parameters described here follows the framework developed and used in \citet{Greif_2019,Raaijmakers_2019,Raaijmakers_2020,Raaijmakers_2021}. Bayes' theorem states that the posterior distribution of  $\boldsymbol{\theta}$ and central energy densities $\varepsilon$ is proportional to the product of the prior distribution of $\boldsymbol{\theta}$, $\varepsilon$ and the nuisance-marginalized likelihood function \footnote{For more discussion of nuisance parameters in this context, see \citet{Raaijmakers_2019}.} 
\begin{equation}
 p(\boldsymbol{\theta}, \varepsilon \mid \boldsymbol{d}, \mathcal{M}) \propto p(\boldsymbol{\theta} \mid \mathcal{M}) p(\varepsilon \mid \boldsymbol{\theta}, \mathcal{M}) p(\boldsymbol{d} \mid \boldsymbol{\theta}, \mathcal{M}) .
\label{like}
\end{equation}
where $\boldsymbol{\theta}$ is the 7-dimensional vector of the EOS model parameters (see Section \ref{subsec:EOS-prior}), $\mathcal{M}$ in this equation denotes the model and $\boldsymbol{d}$ is the dataset. In this work, weighted sampling of the parameter vector $\boldsymbol{\theta}$ is accomplished by the nested sampling Monte Carlo algorithm MLFriends \citep{2016S&C....26..383B,2019PASP..131j8005B} using the UltraNest\footnote{\url{https://johannesbuchner.github.io/UltraNest/}} package \citep{2021JOSS....6.3001B}.\footnote{3000-5000 live points were utilized, depending on the level of complexity of the obtained posterior samples. The Slice sampler in UltraNest was employed, which is well-suited and efficient for high-dimensional sampling. It also ensures consistency in the convergence speed of the sampling process.}

For the {\it current constraints}, we consider the following astrophysical inputs: the most recent mass reported for the heavy pulsar PSR J0740+6620 derived from radio timing \citep[$2.08 \pm 0.07$ \msol,][]{Fonseca_2021}; the tidal deformabilities for the neutron star binary mergers GW170817 and GW190425 reported by the LIGO Scientific Collaboration \citep{LIGOScientific:2017vwq,GW190425}; and the masses and radii inferred from the NICER observations of PSR J0030+0451 and PSR J0740+6620 by \citet{Riley2019} and \citet{Riley2021}. Given that all of the measurements are independent, we can rewrite the likelihood function:
\begin{equation}
\begin{aligned}
 &p(\boldsymbol{\theta}, \varepsilon \mid \boldsymbol{d}, \mathcal{M}) \propto p(\boldsymbol{\theta} \mid \mathcal{M}) p(\varepsilon \mid \boldsymbol{\theta}, \mathcal{M})\\
&\times \prod_{i} p\left(\Lambda_{1, i}, \Lambda_{2, i}, M_{1, i}, M_{2, i} \mid d_{\mathrm{GW}, \mathrm{i}}\left(, \boldsymbol{d}_{\mathrm{EM}, \mathrm{i}}\right)\right)\\
&\times \prod_{j} p\left(M_{j}, R_{j} \mid d_{\mathrm{NICER}, \mathrm{j}}\right)\\
&\times \prod_{k} p\left(M_{k} \mid \boldsymbol{d}_{\mathrm{radio, \textrm {k }}}\right).
\end{aligned}
\end{equation}
Note that we equate the nuisance-marginalized likelihoods to the nuisance-marginalized posterior distributions \citep[for more details of this step and why it is justified see Section 2 of][]{Raaijmakers_2021}.
           
When treating the GW events we fix the chirp mass $M_{\text {c }}=\left(M_{1} M_{2}\right)^{3 / 5}/\left(M_{1}+M_{2}\right)^{1 / 5}$ to the median value $M_{\text {c1}}=1.186$ \msol~for GW170817 and $M_{\text {c2}}=1.44$ \msol~for GW190425. It was shown in \citet{Raaijmakers_2021} that the small bandwidth of the chirp masses has almost no significant influence on the posterior distribution, contributing less than the sampling noise.  We therefore fix the chirp mass, which is beneficial in also reducing the dimensionality of the parameter space and hence the computational cost.  
To speed up convergence of our inference process, we transform the GW posterior distributions to include the two tidal deformabilities, chirp mass and mass ratio $q$, simultaneously reweighing such that the prior distribution on these parameters is uniform. The posterior then becomes
\begin{equation}
\begin{aligned}
&p(\boldsymbol{\theta}, \varepsilon \mid \boldsymbol{d}, \mathcal{M}) \propto p(\boldsymbol{\theta} \mid \mathcal{M}) p(\varepsilon \mid \boldsymbol{\theta}, \mathcal{M}) \\
&\quad \times \prod_{i} p\left(\Lambda_{1, i}, \Lambda_{2, i}, q_{i} \mid \mathcal{M}_{c}, \boldsymbol{d}_{\mathrm{GW}, \mathrm{i}}\left(, \boldsymbol{d}_{\mathrm{EM}, \mathrm{i}}\right)\right) \\
&\quad \times \prod_{j} p\left(M_{j}, R_{j} \mid \boldsymbol{d}_{\mathrm{NICER}, \mathrm{j}}\right) \\
&\quad \times \prod_{L} p\left(M_{k} \mid \boldsymbol{d}_{\text {radio } \mathrm{k}}\right) .
\end{aligned}
\end{equation}
where $\Lambda_{2,i}=\Lambda_{2,i}(\boldsymbol{\theta} ; q_{i})$ is the tidal deformability. We follow the same convention as in \cite{2018PhRvL.121p1101A} and define $M_{1}>M_{2}$, since the gravitational wave likelihood function is degenerate under exchange of the binary components.
\begin{table*}
\begin{tabular}{ccccccccccc}
\hline \hline \text { Model } & $m_{\sigma}$ & $m_{\omega}$ & $m_{\rho}$ & $g_{\sigma }^{2}$ & $g_{\omega }^{2}$ & $g_{\rho }^{2}$ & $\kappa$ & $\lambda$ & $\zeta$ & $\Lambda_{\omega}$ \\
& $(\mathrm{MeV}) $& $(\mathrm{MeV})$ & $(\mathrm{MeV})$ & & & & & & & \\
\hline 
FSU2R  & 497.479 & 782.500 & 763.000 & 107.58 & 182.39 & 206.43 & 3.0911 & -0.001680 & 0.024 & 0.045 \\
TM1-2$\omega\rho$  & 511.198 & 783.000 & 770.000 & 99.97 & 156.34 & 127.75 & 3.5235 & -0.004739 & 0.012 & 0.030 \\
BMPF260  & 500.000 & 782.500 & 763.000 & 100.02 & 161.82 & 100.70 & 2.3030 & -0.017016 & 0.002 & 0.038 \\
\hline
Posterior  & 497.479 & 782.500 & 763.000 & 108.22$^{+7.49}_{-7.32}$ & 175.67$^{+21.42}_{-17.64}$ & 140.05$^{+47.15}_{-44.72}$ & 2.47$^{+1.34}_{-1.37}$ & 0.0035$^{+0.0095}_{-0.0092}$ & 0.0228$^{+0.0047}_{-0.0048}$ & 0.0225$^{+0.0153}_{-0.0154}$ \\
\hline \hline
\end{tabular}
\caption{This table shows the FSU2R, TM1-2$\omega\rho$ and BMPF260 parameter vectors (the latter two being used to generate simulated M-R measurements to test parameter recovery). The row labelled Posterior gives the median and 68\% credible intervals for the parameters as inferred from current observations, see Section \ref{results_now}.
}

\label{Set-up}
\end{table*}
        
For the {\it future constraints}, we consider the dataset $\boldsymbol{d}$ to be composed of  M-R constraints of the quality that we anticipate from the next-generation X-ray telescopes (such as STROBE-X or eXTP). This is relatively straightforward to predict given that uncertainties in masses and radii should scale in a simple fashion with exposure time and telescope effective area \citep{Lo2013,Psaltis2014}.  While new GW measurements of tidal deformabilities are anticipated on a similar timescale, it is hard to predict the quality of these given the dependence on source properties and the uncertainties in merger rates; it is also difficult to say with certainty whether we can expect an improved or increased maximum mass measurement from radio timing. We thus do not include any future GW or radio constraints in our {\it future constraints} simulations. It is also valuable to consider what can be achieved by a single technique, for the purposes of independent cross-checks of different methods.   

In this scenario, with only inferred masses and radii from X-ray pulse profile modelling, the likelihood function in Eq.~(\ref{like}) is given by:
\begin{equation}
\begin{aligned}
&p(\boldsymbol{\theta}, \varepsilon \mid \boldsymbol{d}, \mathcal{M}) \propto p(\boldsymbol{\theta} \mid \mathcal{M}) p(\varepsilon \mid \boldsymbol{\theta}, \mathcal{M}) \\
&\times \prod_{j} p\left(M_{j}, R_{j} \mid d_{\mathrm{Future/Future-X}, j}\right) .
\end{aligned}
\end{equation}

Based on studies of the capabilities of future telescopes \citep{Watts_2016,Watts2019} we define two Future observation cases for study (similar to the scenarios considered in \citealt{Rutherford22}). In our "Future" set-up, somewhat beyond what is expected to be achievable with NICER, we assume that we will have six M-R measurements which we model as bivariate Gaussians with 5\% uncertainty, distributed from $\sim$ 1.2 \msol~to $\sim$ 2.2~\msol, which are centered at [1,2, 1.4, 1.9, 2.0, 2.1, 2.2] \msol. These simulated measurements span a reasonable range compared to current observations and include three values corresponding to known masses for current NICER sources: PSR J0740+6620 (2.1 \msol, \citealt{Cromartie2020,Fonseca_2021}), PSR J1614-2230 (1.9 \msol, \citealt{Demorest2010}) and PSR J0437-4715 (1.4 \msol, \citealt{Reardon2016}).  In the "Future-X" scenario we upgrade our precision of the previous six measurements to the 2\% uncertainty level, distributed over the same mass range. This represents a `best case' scenario for what we might be able to achieve with a telescope like STROBE-X or eXTP.  

\begin{figure*}
\begin{multicols}{2}
    \includegraphics[width=\linewidth]{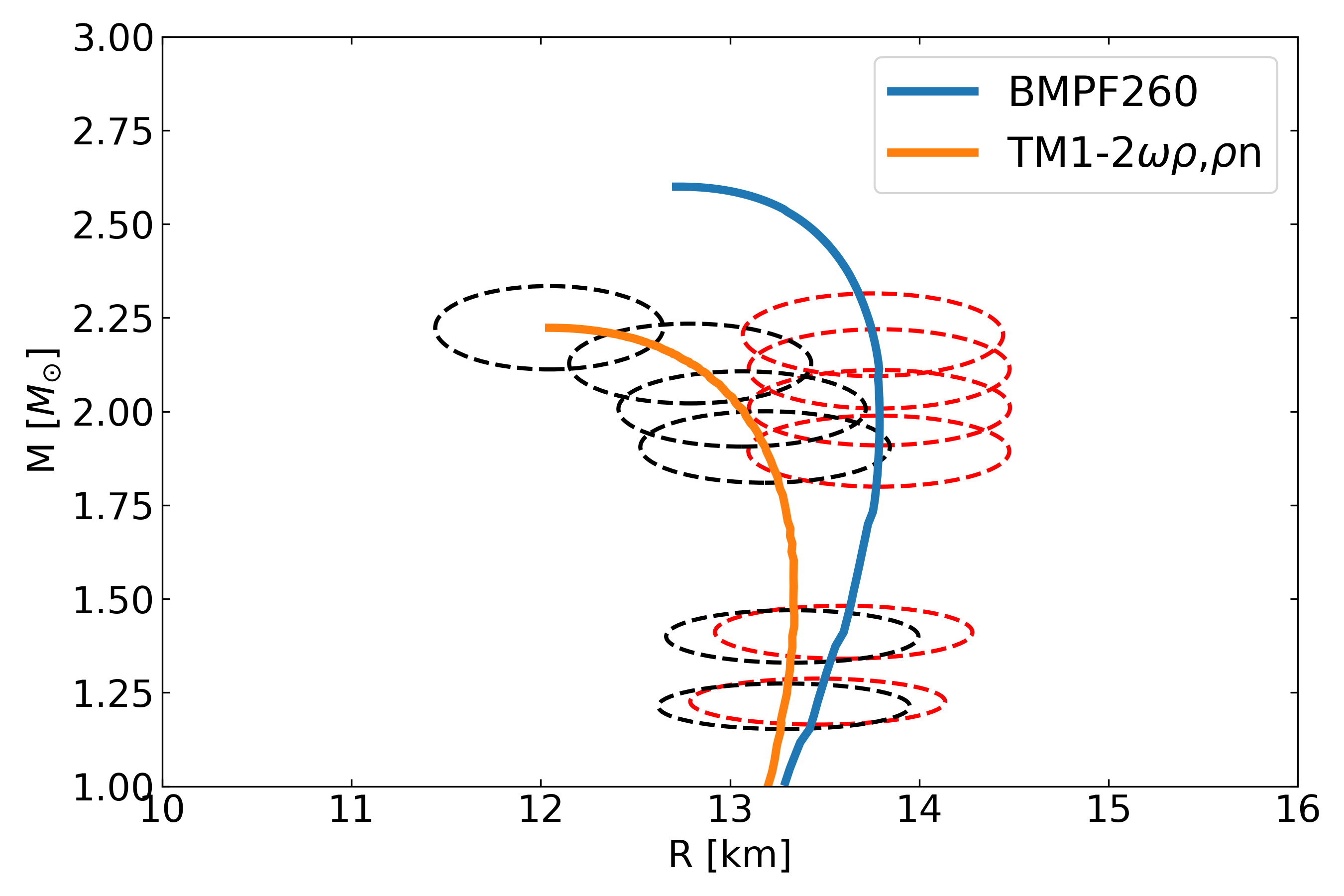}\par 
    \includegraphics[width=\linewidth]{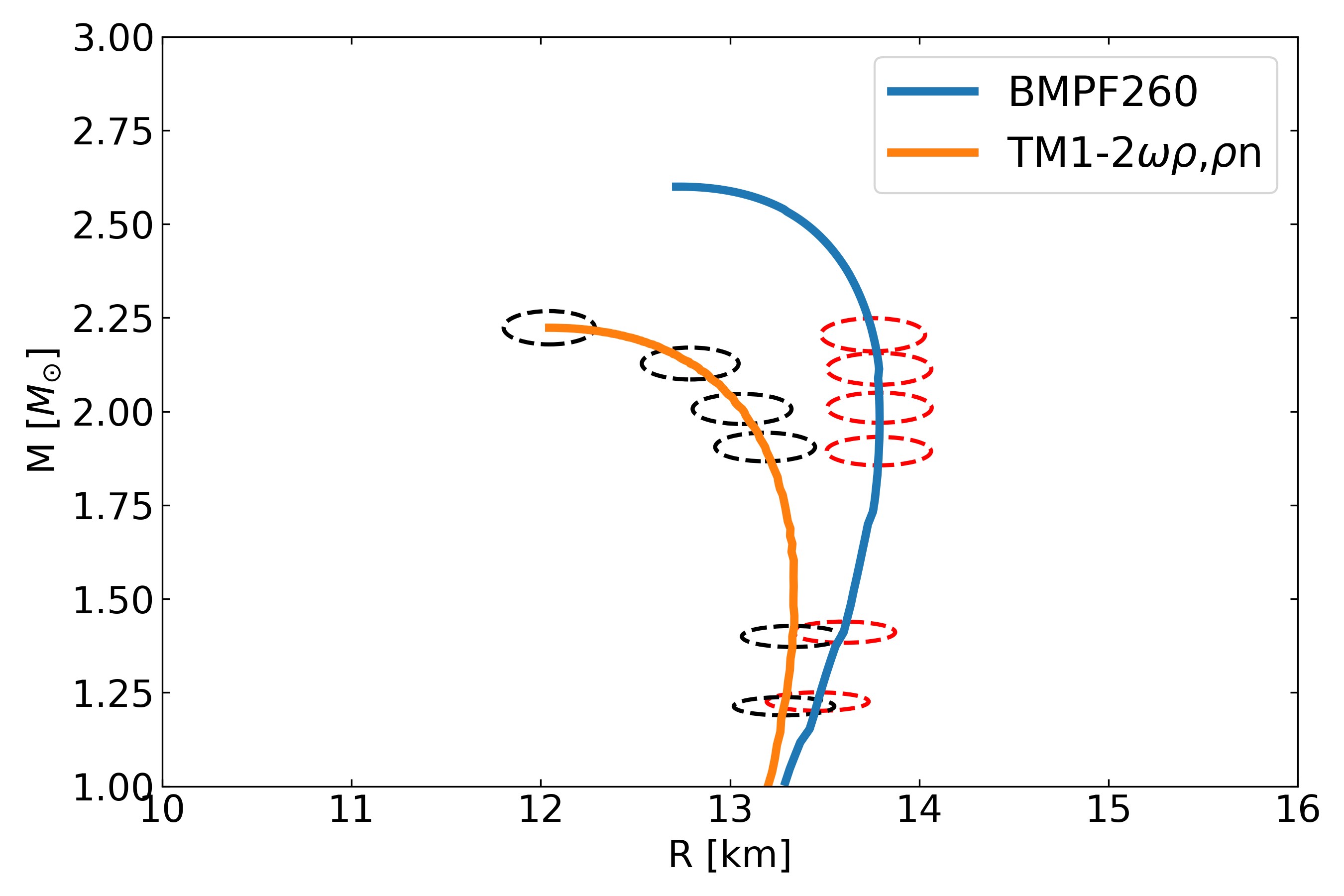}\par 
    \end{multicols}
\caption{M-R curves for the TM1-2$\omega\rho$ EOS and BMPF260 EOS, which are two underlying scenarios (injected EOS parameter vectors) used to generate simulated Mass Radius measurements for the {\it future constraints} scenario. The red dashed curves are the six simulated M-R posteriors, which are centered at [1.2, 1.4, 1.9, 2.0, 2.1, 2.2] \msol.  Left panel: corresponding to {\it Future} scenario with six 5\% uncertainty M-R observations, Right panel: corresponding to {\it Future-X} scenario with six 2\% uncertainty M-R posteriors}
\label{fig:ground_truth}
\end{figure*}

To assess parameter recovery, we now select some specific EOS parameter vectors to test.  Picking such a vector determines the M-R relation that is used to generate simulated M-R measurements (for the mass vector given in the previous paragraph). 
We choose the TM1-2$\omega\rho$, $pn$ EOS model \citep{Providencia:2012rx} and the BMPF260 model \citep{Malik:2023mnx}, which are based on the same underlying model as FSU2R. Their parameter vectors are given in Table.\ref{Set-up}, with the parameters for FSU2R shown for comparison.  The resulting M-R relations. together with the simulated measurements, are shown in Figure \ref{fig:ground_truth}. Note that in this study, due to computational constraints, we consider only two injected parameter vectors; while our results are still illustrative of the capabilities of future missions, follow-on studies should ideally examine parameter recovery for a broader range of injected models (and simulated M-R posteriors).
          
\section{EOS constraints from current observations}
\label{results_now}

 In this section we investigate how well existing observations constrain both the EOS parameters and the corresponding nuclear matter saturation properties. We use the masses and radii inferred from NICER data by \citet{Riley2019} for the pulsar PSR J0030+0451 ($M= 1.34_{-0.16}^{+0.15}$ \msol~and $R = 12.71_{-1.19}^{+1.14}$ km) and by \citet{Riley2021} for the heavy pulsar PSR J0740+6620 ($M= 2.07\pm 0.07 $ \msol~and $R = 12.39_{-0.98}^{+1.30}$ km), and the two GW tidal deformability measurements \citep{LIGOScientific:2017vwq,GW190425}, with GW170817 in particular favouring softer EOS.  

\begin{figure*}
	\centering
	\includegraphics[scale=0.4]{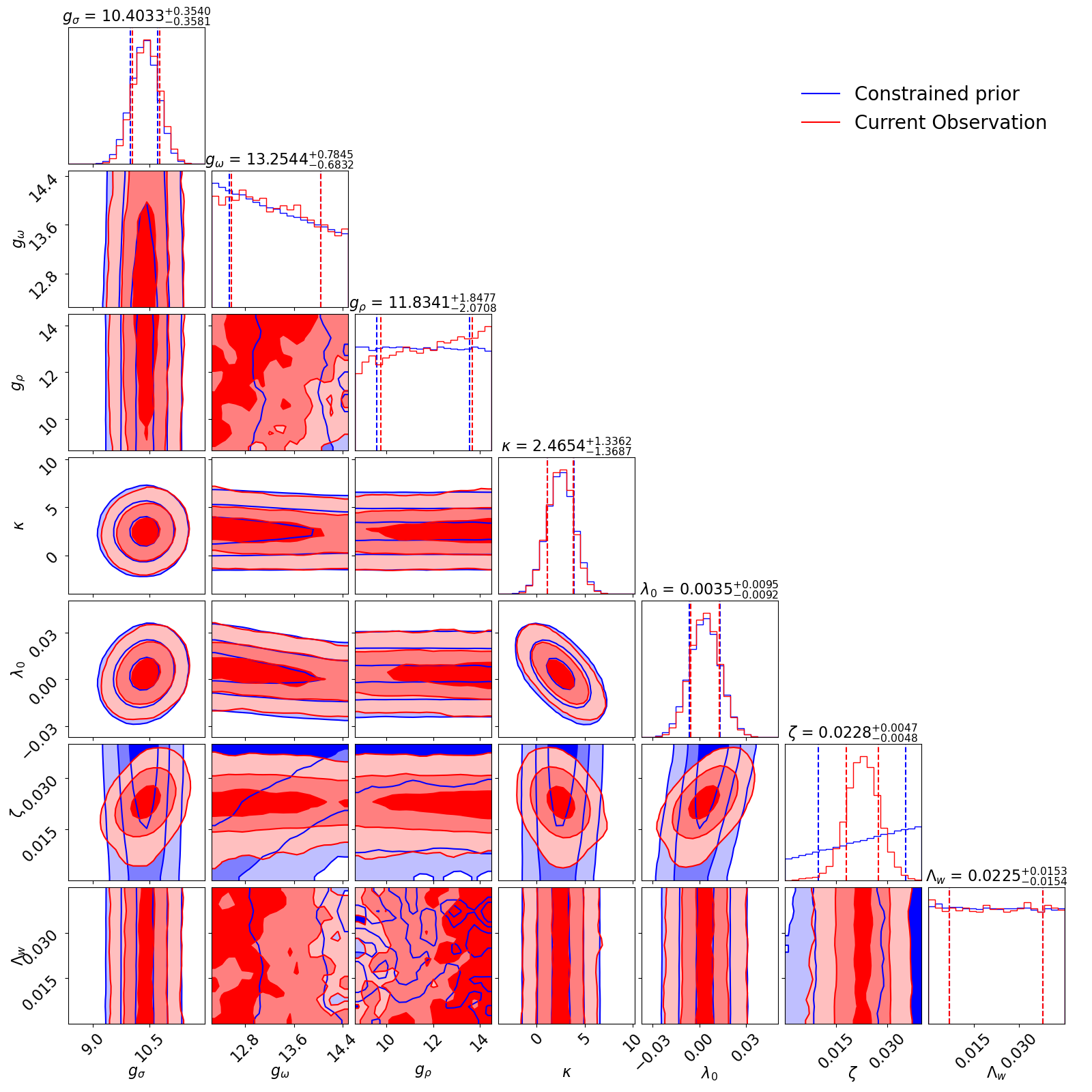}
	\caption{The posterior of the nucleonic EOS model parameters after applying constraints for all existing observations. Blue indicates the constrained priors, red the posteriors. The contour levels in the corner plot, going from deep to light colors, correspond to the 68\%, 84\%, and 98.9\% levels. The dashed line in the 1D corner plots represents the 68\% credible interval, and the title of that plot indicates the median value of the distribution as well as the range of 68\% credible interval. Here $\kappa$ is given in MeV.
	}
	\label{J0740}
\end{figure*} 

 In Figure \ref{J0740} we show the posterior distribution of the EOS parameters. The only parameters for which we see variation from the prior are $\zeta$, $g_{\rho}$ and a very small shift in $g_\omega$.   This result is reasonable considering the precision of the current measurements: it is still not possible to extract strong constraints on all of the model parameters from the current M-R and tidal deformability measurements.  

\begin{figure}
	\centering
	\includegraphics[width = \columnwidth]{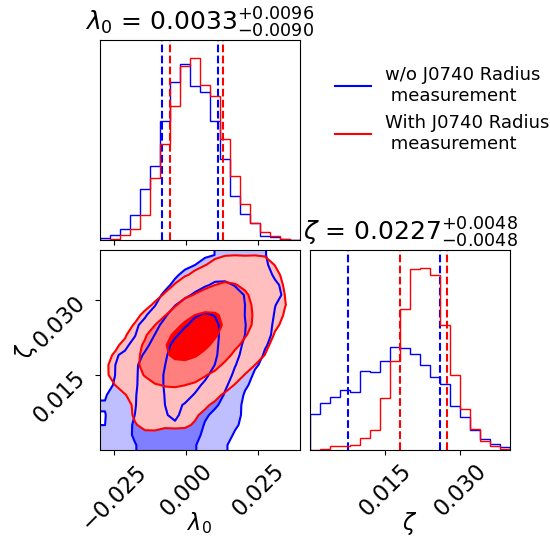}
	\caption{Comparison between the observational constraint with (red) and without (blue) the NICER PSR J0740+6620 radius measurement. The contour levels in the corner plot, going from deep to light colors, correspond to the 68\%, 84\%, and 98.9\% levels. The dashed line in the 1D corner plots represents the 68\% credible interval, and the title of each plot indicates the median value of the distribution as well as the range of 68\% credible interval.
	}
	\label{NICER_compare}
\end{figure}

\begin{figure}
	\centering
	\includegraphics[width = \columnwidth]{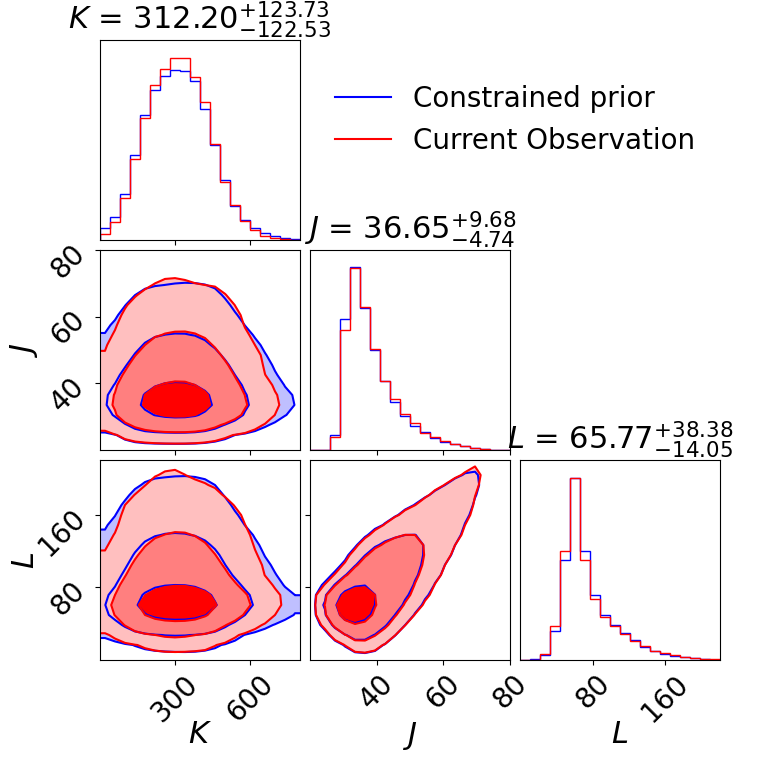}
	\caption{The posterior of all the nuclear quantities after applying constraints from current astrophysical observations. The contour levels in the corner plot, going from deep to light colors, correspond to the 68\%, 84\%, and 98.9\% levels. The dashed line in the 1D corner plots represents the 68\% range, and the title of that plot indicates the median value of the distribution as well as the range of 68\% credible interval.
	}
	\label{NICER_nuclear_quantities}
\end{figure}

Significant constraints are achieved for $\zeta$, a parameter which influences both the maximum mass stellar mass and radius simultaneously: both of them increase when $\zeta$ decreases. Extreme values are disfavoured; instead the data favour a middle value of $\zeta$ that allows a maximum mass compatible with PSR J0740+6620 and radii that are consistent with both the NICER and GW measurements.  Fig.\ref{NICER_compare} illustrates the constraints that would have been delivered had we used only the mass measurement for PSR J0740+6620 from \citet{Fonseca_2021}, rather than the NICER M-R constraint for that source. The posterior on $\lambda_{0}$ shifts very slightly due to its sensitivity to the radius. However the $\zeta$ parameter distribution becomes much sharper once the radius information is included, and values of $\zeta < 0.01$ become highly disfavoured. The parameters $\lambda_{0}$ and $\zeta$ affect the symmetric nuclear matter properties, and, therefore, the incompressibility $K$ is the nuclear matter property most affected by {\it current observation} constraints as seen in Figure \ref{NICER_nuclear_quantities}, where the posterior of all the nuclear quantities after applying constraints
from current astrophysical observations are shown. Notice that in our analysis the other symmetric nuclear matter properties ($E/A$, $m^*$ and $n_0$) are allowed to vary in a small range.  The $g_\rho$ shows a small increase which results in an slight increase of the symmetry energy at saturation, however, these changes are not very significant (see Figure \ref{NICER_nuclear_quantities}).
 
\begin{table}
\begin{tabular}{lcc}
\hline \hline \text { Model}  & \text{ln(Z)} & \text{Bayes' factor ($Z/Z_\mathrm{Full~RMF}$)}\\
\hline 
\text{FSUGarnet} & -2.810 &   7.83     \\

\text{FSU2}  &  -2.942 & 6.86     \\

\text{FSU2R}  & -2.967 &  6.69    \\

\text{TM1-2$\omega\rho$} & -3.795 &  2.92  \\

\text{BigApple}  & -5.397 &  0.589  \\

\text{NL3}   & -6.986 &    0.120        \\

\text{IU-FSU}    &  -7.356   & 0.083     \\

\hline
\text{RMF: Full model} & -4.868 &  $\cdots$ \\

\text{RMF: Max posterior}  & -2.753 & 8.29 \\
\hline \hline
\end{tabular}
\caption{This table gives the global log evidence ($\ln Z$), as returned by Ultranest, for various different EOS models and current astrophysical constraints.  For individual EOS models (above the line) the EOS model parameters are fixed, and the only free parameters in the sampling are the central densities of the stars for which we have constraints.  For our full parameterized RMF model (below the line) the EOS parameters are free as well; this affects the evidence computation since the prior space is larger.  To illustrate this, we also give $\ln Z$ for the case where we fix the RMF EOS parameters to those of the maximum posterior sample in the full analysis; in this run only the central densities vary.  In this sense it is more comparable to the other individual models. }
\label{EOS_ev}
\end{table}

One approach to evaluate the plausibility of the EOS models based on the same framework used in this work is to compute the Bayesian evidence for that model given current measurements.  Table \ref{EOS_ev} gives the global log evidence ($\ln Z$) as returned by Ultranest for both individual EOS models with fixed parameters and our full RMF model.  For the fixed EOS parameter models the only free parameters in the sampling process are the central densities of the stars for which astrophysical constraints are available.  For the full RMF model the EOS parameters are also free and this affects the global evidence calculation due to the larger prior space. To illustrate this we also give $\ln Z$ for a run where we fix the EOS parameters of the RMF model to those of the maximum posterior sample, so that only the central densities vary.  As might be expected, the evidence improves by an amount that should be borne in mind when comparing to the other fixed parameter models. 

In comparing the models we use Bayes' factors.   \citet{Kass} deem a model `substantially preferred' if the Bayes' factor is more than 3.2 and 'strongly preferred' if more than 10. By this metric the full RMF model is strongly preferred compared to IU-FSU, and substantially preferred compared to NL3.  FSUGarnet, FSU2 and FSU2R are substantially preferred compared to the full RMF model (but are statistically indistinguishable if one considers instead the Bayes' factor computed with reference to the Max Posterior RMF run).

\begin{figure}
	\centering
	\includegraphics[width = \columnwidth]{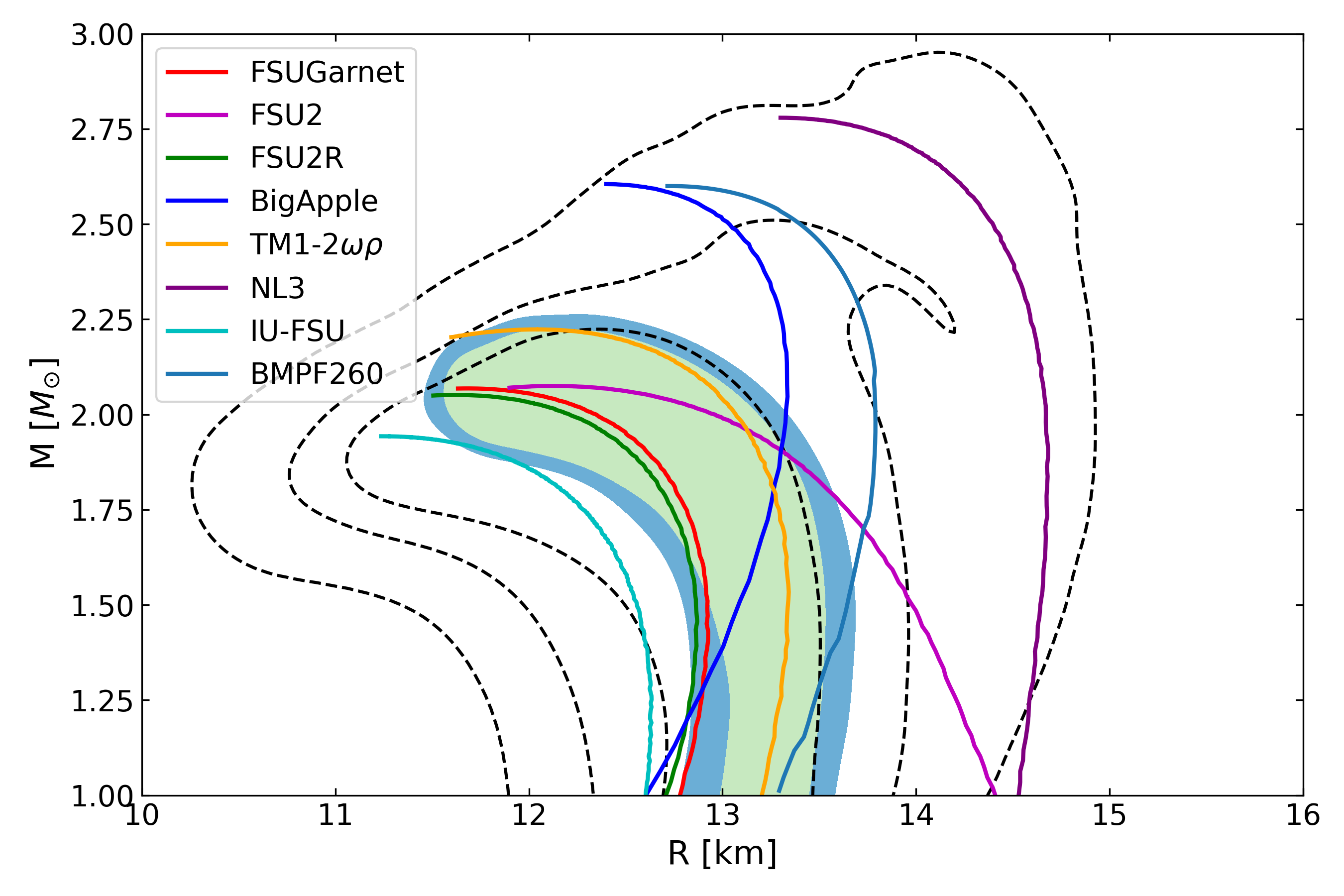}
	\caption{The posterior in M-R space once all current astrophysical constraints are taken into account, showing the 68\% (light green) and 84\% (blue) credible regions. For comparison we show M-R relations for various individual EOS models which map to specific individual parameter vectors within the broad prior space of our model. The outer black dashed line delineates the shape of our constrained M-R prior, delineating the 68\%, 84\% credible region and 100\% credible regions (as in Figure \ref{M-R_nu}).
	}
	\label{NICER-M-R}
\end{figure}

In Figure \ref{NICER-M-R}, we present the M-R posterior generated as a result of incorporating all current observational constraints. For the purpose of comparison, we have also plotted the EOS models that have been evaluated for evidence. From this figure, it is evident that the evidences of the EOS are correlated with their relative positioning to the inferred M-R posterior contour. The maximum posterior EOS set-up (denoted as "MAX", detailed parameterization of this EOS in Table \ref{Set-up}) has been plotted. It is seen that it does not fully traverse the region of maximum probability in the M-R space. This is understandable as the parameters utilized in this set-up essentially affect the high-density EOS. As a result, the MAX method is able to traverse a wide range of the highest-probability M-R posterior having high central energies, but is limited in its ability to traverse the low mass high posterior probability in the M-R space, approximately below 1.4 \msol. This region is more closely influenced by the EOS of the inner crust. In future studies, it will be beneficial to utilize a universal EOS set-up that incorporates models for the core, inner crust, and outer crust in order to overcome this limitation. As is evident from the M-R contour presented, current observations, despite having ample room for improvement in terms of precision, are already capable of imposing a significant constraint on the EOS within the corresponding M-R space.

\section{EOS constraints from future mass-radius measurements}
\label{results_future}

In this section we investigate the effectiveness of using simulated M-R measurements, of the type expected from future missions, to constrain the EOS.  We do not consider any tidal deformability measurements, focusing purely on what can be delivered by pulse profile modelling.  We consider two scenarios: six measurements with a 5\% uncertainty on the M-R of a single star, referred to as the Future scenario, and six measurements with a 2\% uncertainty on the M-R of a single star, the Future-X scenario. This section is organized as follows. Results using TM1-2$\omega\rho$ as the injected model are presented in Section \ref{future_TM1-2} for the Future scenario and Section \ref{futurex__TM1-2} for Future-X. We compare the two scenarios in Section \ref{compare_futures}. In Section \ref{futures_model2} we repeat this analysis using a different injected model, BMPF260.

\subsection{Constraints from the "Future" scenario}
\label{future_TM1-2}

\begin{figure*}
	\centering
	\includegraphics[scale=0.4]{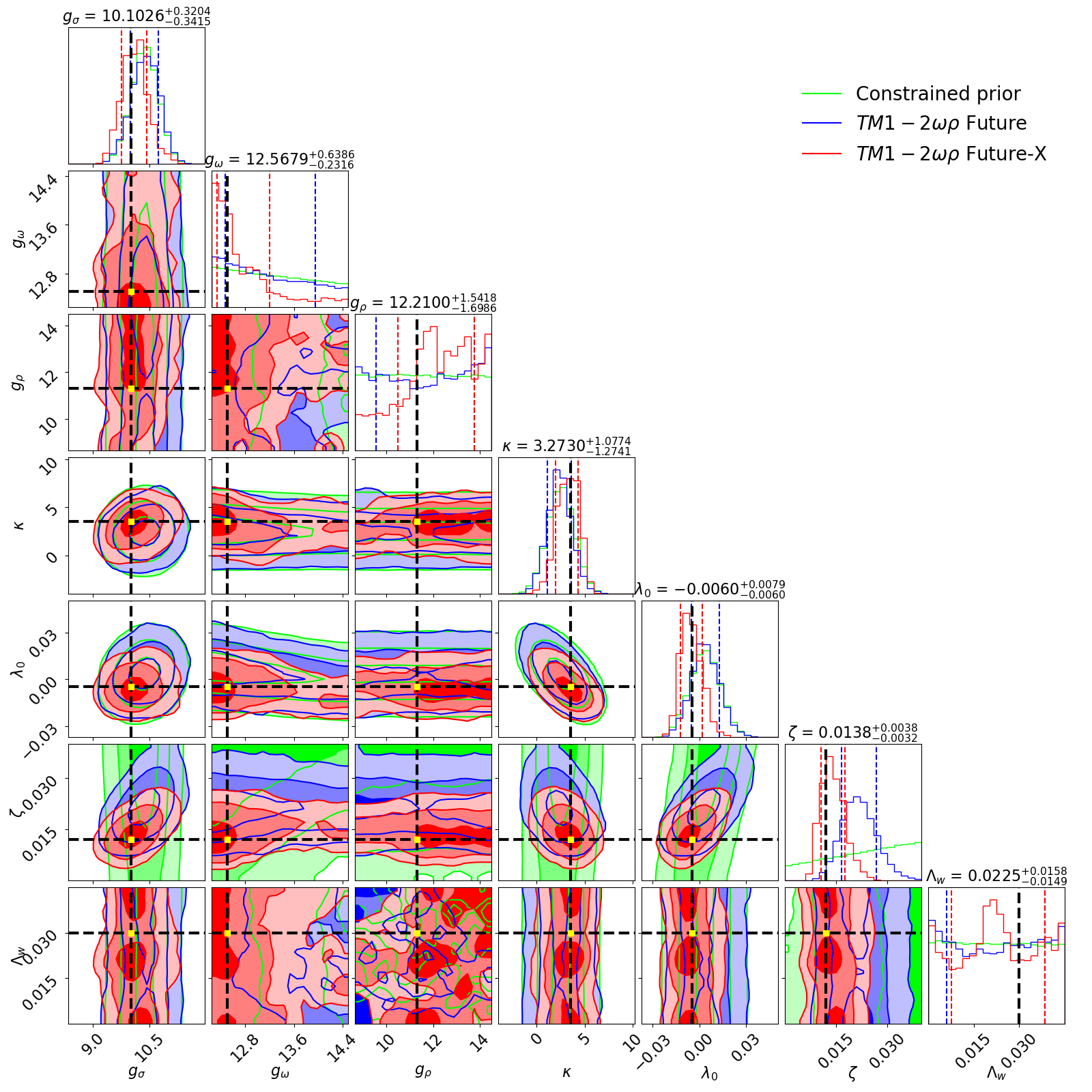}
	\caption{Comparison between the observational constraints from the Future and Future-X scenarios using TM1-2$\omega\rho$ as the injected EOS model used to generate the synthetic M-R posteriors.  Lime green is the constrained prior, blue is the Future posterior and Red the Future-X posterior. The contour levels in the corner plot, going from deep to light colors, correspond to the 68\%, 84\%, and 98.9\% levels. The dashed line in the 1D corner plots represents the 68\% range, and the title of that plot indicates the median value of the distribution as well as the range of 68\% credible interval. Here $\kappa$ in MeV. The black dashed horizontal and vertical lines in the plot and yellow dots show the injected values used to generate the simulated M-R measurements.
	}
	\label{pro_mini}
\end{figure*}

In this section we investigate the impact of the Future scenario on the constraints for the nucleonic model. The simulated M-R measurements are shown in the left panel of Figure \ref{fig:ground_truth}. The posterior of the EOS parameters is compared to the constrained priors in Figure \ref{pro_mini}. 

With the implementation of more stringent constraints, 
the posterior distribution of $g_{\omega}$ shifts a little towards lower values, serving as a potential screening tool for various EOS models that incorporate this parameter.  The distributions of $g_{\rho}$ and $\Lambda_{\omega}$ shift only very slightly to favor two extremities.

The parameters $\kappa$ and  $\lambda_0$ are constrained a little more tightly than the prior, with a slightly narrower range favored by the inference.  The parameter $\zeta$ is reshaped into a Gaussian-like distribution favoring a median value, with most of the prior space being excluded.

\begin{figure*}
\begin{multicols}{2}
    \includegraphics[width=\linewidth]{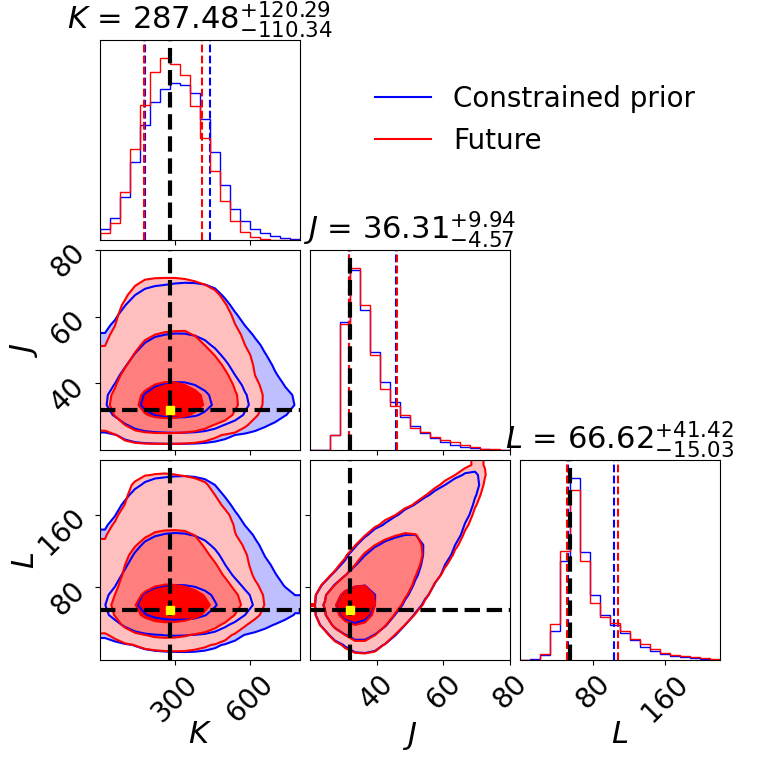}\par 
    \includegraphics[width=\linewidth]{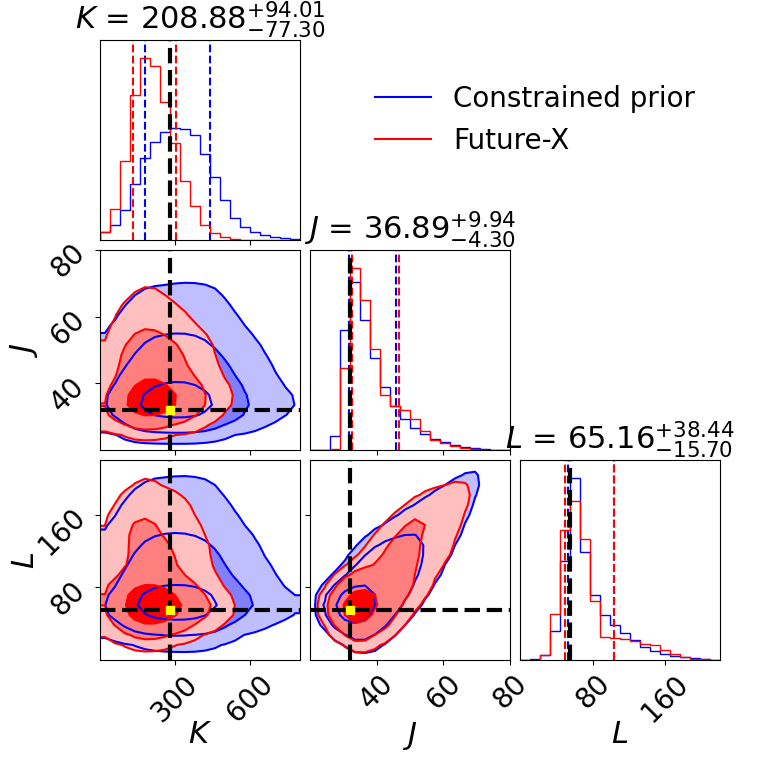}\par 
    \end{multicols}
\caption{The posterior distributions of the nuclear quantities in the Future (left) and Future-X (right) scenarios for TM1-2$\omega\rho$ as the injected model. In both panels, blue shows the constrained prior.  The contour levels in the corner plots, going from deep to light colors, correspond to the 68\%, 84\%, and 98.9\% levels. The dashed line in the 1D corner plots represents the 68\% range, and the title of that plot indicates the median value of the distribution as well as the range of 68\% credible interval. The black dashed horizontal and vertical lines in the plot and yellow dots show the injected values used to generate the simulated M-R measurements. 
}
\label{fig:TM1-2}
\end{figure*}

One notable observation is that even though we implemented more precise measurements compared to current observations, the exclusion of parameter space is only slightly better (comparing Figure \ref{J0740} and Figure \ref{pro_mini}).  However note that this is now delivered by M-R measurements alone, with no GW input - an important step to facilitating cross-comparisons of the different techniques and any potential systematic or modelling errors.  The nuclear quantities resulting from the Future posterior are seen in left panel of Figure \ref{fig:TM1-2}. Interestingly the incompressibility $K$ shows stronger constraints from the simulated measurements (Figure \ref{fig:TM1-2}), compared to the {\it current constraints} (Figure \ref{NICER_nuclear_quantities}). This underscores the importance of having radius measurements of high mass stars. 

\subsection{Constraints from the "Future-X" scenario}
\label{futurex__TM1-2}

In this section we explore the constraints arising from the Future-X scenario, in which the previous six simulated measurements now have 2\% M-R uncertainty. The simulated M-R measurements are shown in the right panel of Figure \ref{fig:ground_truth}.
In Figure \ref{pro_mini}, the posterior distribution of the EOS parameters is compared to the prior. 

The Future-X scenario, unlike the previous analyses, demonstrates the capability to extract robust constraints for almost all of the EOS parameters. This results in a significant decoupling of the parameters from the prior distribution and reflects the fundamental properties of neutron stars. The parameters serve as a set of reliable indicators for distinguishing between numerous EOS, as demonstrated in Table \ref{EOS_ev}. This provides stronger evidence for the exclusion or favoring of different models based on the posterior obtained from Future-X.

In Fig.~\ref{pro_mini} the constraint on $\zeta$ becomes even tighter than before. More than half of the $g_{\omega}$ prior space falls outside of the 68\% Future-X range, as the inference homes in on the input value. The inferred value of $g_\sigma$ also shifts slightly: note that there is a link between $g_{\sigma}$ and $g_{\omega}$ to reproduce the correct binding energy. The distribution of $g_{\rho}$ generally favors a higher value, significantly reducing the 68\% credible interval. A shift in the distribution of $\kappa$ and $\lambda_0$ can also be observed. The behavior of the posterior distribution of $\Lambda_{\omega}$ is particularly interesting. While no region of parameter space is excluded, we have seen the posterior distribution of this parameter starts to show some structure, meaning that it has been constrained. This trend was not observed in the Future scenario posterior. Note that $\Lambda_{\omega}$ is directly related to the symmetry energy and it is expected that this parameter is mostly sensitive to observations that depend on the composition - in particular the proton fraction - such as cooling.

The nuclear quantities inferred from the Future-X posterior are depicted in the right panel of Figure \ref{fig:TM1-2}. It is noteworthy that the posterior distribution of the incompressibility $K$ has shifted significantly and a considerable amount of parameter space has been excluded. However this remains the only derived nuclear quantity that is strongly constrained by these simulated observations. 

\subsection{Comparing Future and Future-X}
\label{compare_futures}

\begin{table}
\begin{tabular}{ccccccccccc}
\hline \hline \text {Parameters}  & Current constraints & Future & Future-X\\
\hline 
$g_{\sigma }$     &0.072620 & 0.099102 & 0.101921\\

$g_{\omega }$     & 0.005315 & 0.039137 & 0.114267\\
$g_{\rho }$      & 0.022659 & 0.117450 &  0.223765\\

$\kappa$       & 0.001402 & 0.045308 & 0.054384\\
$\lambda_0$    & 0.038278 & 0.061046 & 0.095424 \\
$\zeta$     & 0.038472 & 0.073276 & 0.073910\\
$\Lambda_{\omega}$  & 0.023653  & 0.034307& 0.037426\\
\hline \hline
\end{tabular}
\caption{This table shows the KL divergence comparison between current observations, Future and Future-X.}
\label{KL-div}
\end{table}	 

Although the range of parameter space excluded may not seem large, particularly for the Future scenario, the constraining power of the measurements in both future scenarios is an improvement over current constraints. To demonstrate this we compute the Kullback-Leibler (KL) divergence, a measure of the parameter-by-parameter information gain of the posterior over - in this case - the constrained prior.  If the KL divergence is zero it implies no information gain; KL divergence is normalized (using the factor $1-\exp(-D_{KL})$, where $D_{KL}$ is the unormalized KL divergence) such that the maximum value is 1. The KL divergence values for the seven parameters are given in Tab. \ref{KL-div}, comparing the Future and Future-X with current observations.  Every parameter shows an improvement as measured by the KL divergence value. 

From this perspective, the Future and Future-X scenarios demonstrate stronger constraining power for some parameters. Our simulated observations constrain some of the parameters that define the symmetric nuclear matter EOS, but are not very sensitive to the symmetry energy, since the parameters $g_\rho$ and $\Lambda_\omega$ are not affected as strongly as the other parameters. 

Concerning the symmetric nuclear matter behaviour, $g_\sigma$ and $g_\omega$  define the binding energy at saturation, $g_\omega$, $\kappa$ and $\lambda_0$  are strongly related with the incompressibility, and 
 $\zeta$ softens the EOS at large densities allowing the reproduction of the mass and radius of the pulsar PSR J0740+6620. However, with the advancement in precision and an increase in the number of measurements, we have shown that the above limitations may be gradually overcome. As precision improves and new measurements are introduced, certain parameters will start to be constrained, decoupled from our prior set-up, as evidenced by a shift in posterior probability or exclusion of a certain parameter space.

The results shown in Figure \ref{pro_mini} demonstrate the increased constraining power of both Future and Future-X on the EOS parameters. It is observed that the parameter space of $g_{\sigma}$, $g_{\omega}$, $\kappa$, $\lambda_0$, and $\zeta$ has become narrower due to the improvement in precision of the simulated measurements. Furthermore, both $g_{\rho}$ and $\Lambda_{\omega}$ exhibit a stronger preference for certain values. Upon comparing the contours of the posterior distribution with the position of the injected parameters used to generate the simulated M-R measurements (indicated by the black dashed lines in the plot), it is evident that an improvement in precision results better recovery of the injected parameters. We note that, with the exception of $\Lambda_{\omega}$, all of the injected parameters (yellow dots) fall within the 68\% credible regions.   For $\Lambda_{\omega}$ this is not unexpected, since this parameter is the least sensitive to M-R measurements and hence still strongly determined by our priors.  However, it is reasonable to anticipate that increasing the precision or the number of M-R measurements will eventually shape $\Lambda_{\omega}$ as a distribution peak at the injected value, thereby providing us with the potential to reproduce the full EOS from inference.

Overall, the goal of constraining the entire EOS parameter space has been achieved, as all parameters have deviated or are starting to deviate from the prior distributions, indicating that we would be entering the data-dominated rather than prior-dominated regime.  
\begin{figure*}
	\centering	\includegraphics[scale=0.4]{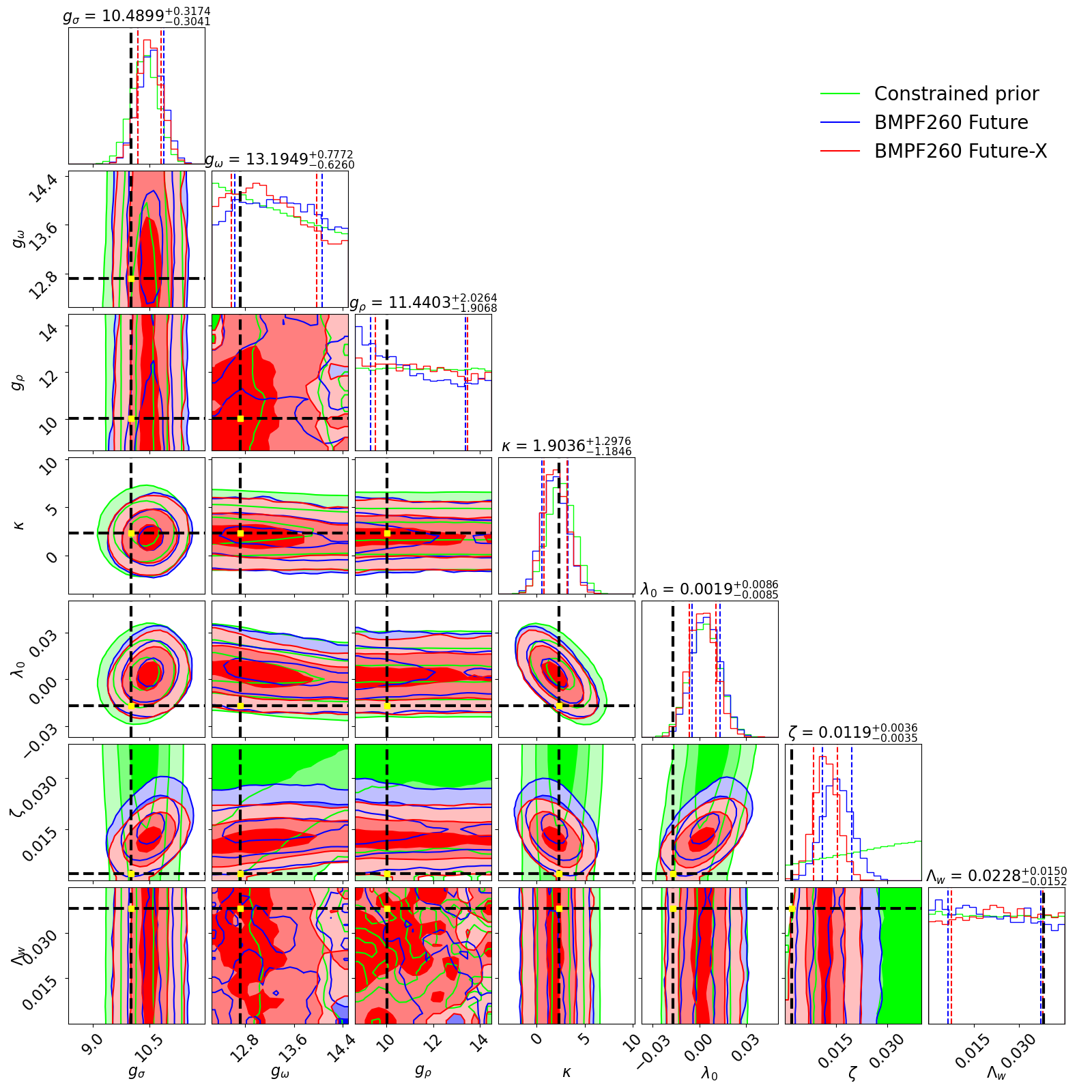}
	\caption{Comparison between the observational constraints from the Future and Future-X scenarios using BMPF260 as the injected EOS model used to generate the synthetic M-R posteriors.  Lime green is the constrained prior, blue is the Future posterior and red the Future-X posterior. The contour levels in the corner plot, going from deep to light colors, correspond to the 68\%, 84\%, and 98.9\% levels. The dashed line in the 1D corner plots represents the 68\% range, and the title of that plot indicates the median value of the distribution as well as the range of 68\% credible interval. Here $\kappa$ in MeV. The black dashed horizontal and vertical lines in the plot and yellow dots show the injected values used to generate the simulated M-R measurements.
	}
	\label{2ground-truth}
\end{figure*} 

\subsection{Comparing different injected parameter vectors}
\label{futures_model2}

In this section we explore the dependence of our results on the injected parameter vector being tested, and study the Future and Future-X scenarios for simulated M-R measurements based on the BMPF260 EOS \citep{Malik:2023mnx}. This particular EOS model allows for stars with a maximum mass of 2.5 \msol and produces larger radii than the  TM$1-2\omega\rho$, for stars with the same mass (see Figure \ref{fig:ground_truth}). By using the same fixed set of stellar masses [1.2, 1.4, 1.9, 2.0, 2.1, 2.2] \msol, we aim to isolate the influence of different radius measurements on the exclusion of parameter space\footnote{Note that in this case, since the injected radius is larger, the absolute uncertainty in radius is higher than for the previous model for our two future scenarios.}.  The results are shown in Figure \ref{2ground-truth}.

For this alternative injected model and this particular choice of injected masses the data are less constraining: the only major improvement is in the constraint on $\zeta$.  Future-X performs better than Future (as we would expect), but even with Future-X we do not recover the injected value of $\zeta$ within the 68\% credible interval.

This illustrates some important issues.  Firstly, it shows that a lower radius for the highest mass star in the data set is - for this model - more constraining.  This can be understood from the M-R prior space shown in Fig.~\ref{M-R_nu} and the influence of $\zeta$. There is simply less prior space that allows lower radii at high mass. It also illustrates however that constraints will be weaker if we are not able to measure the full spread of masses allowed by the underlying EoS: the maximum mass star in our synthetic data set is well below the maximum allowed by the  BMPF260 model. Constraints would be tighter if we had a higher mass star in the sample, and this is one motivation for trying to increase the sample size as much as possible for future instruments. A high mass measurement obtained from a complementary technique (radio pulsar timing, or gravitational wave measurements if the compact object could be unequivocally identified as a neutron star) could also be incorporated.

\section{Additional physics output from current constraints}
\label{physics}
\begin{figure}
	\centering
	\includegraphics[scale=0.41]{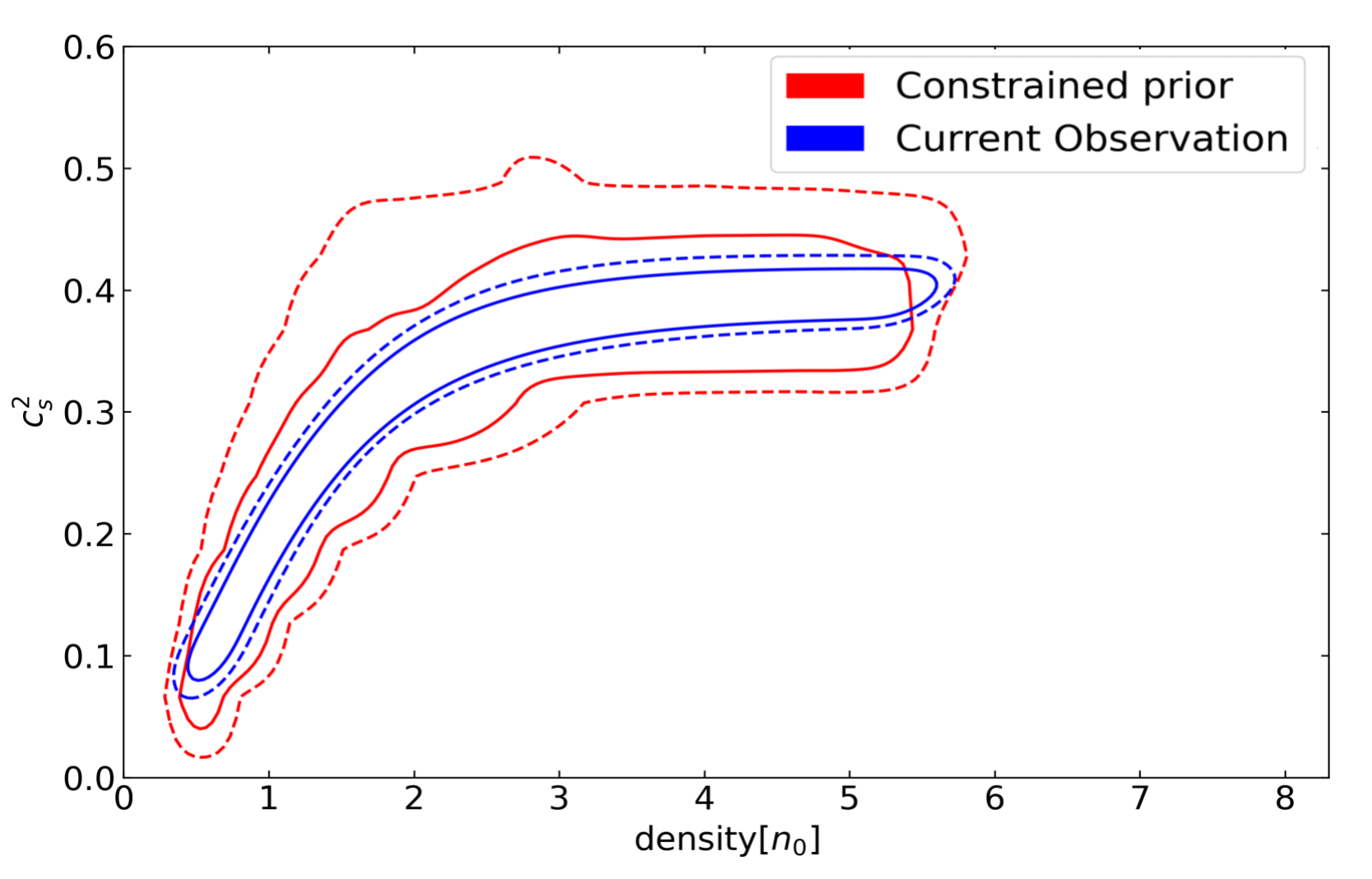}
	\caption{The posterior distribution of $c_{s}^2$ derived from current observations (blue solid line - 68\% credible interval, blue dashed line - 84\% credible interval), compared to the constrained prior distribution (same levels) in red.}
	\label{ss}
\end{figure} 

In addition to the M-R information that can be used to constrain the EOS, more information can be extracted from these inferences. One particularly interesting piece of information is the constraint on the speed of sound from our model, which is computed from the derivative of the pressure with respect to the energy density. This information can be computed by resampling the posterior EOS parameter space. The result can be seen in Figure \ref{ss}, which shows the posterior distribution of $c_{s}^2$ after applying constraints from all available current observations. This figure also shows the constrained prior distribution, which allows us to conclude that the current observations have a constraining power on the speed of sound. This reflects the constraining power previously identified on the isoscalar channel, in particular on the $\zeta$ coupling. As we are using a relativistic mean field theory framework, we expected causality to be automatically equipped as a relativistic model. Our result is surprising because the inferred speed of sound never exceeds $\sqrt{0.45}c$ at the 84\% level, and approaches a stable value after reaching a sufficiently high density. The speed of sound behavior obtained in the present inference analysis is rather different (more restrictive, no turnover at an intermediate density and no extension to values above $\sqrt{0.7}c$) compared to what has been obtained from previous speed of sound inferences using NICER and GW data \citep[see e.g.][]{Raaijmakers_2021, Legred21,Gorda2022,Annala2022,Altiparmak22,Annala23}.
In \cite{Bedaque2015}, the authors have shown that a speed of sound always below the conformal limit would be in tension with the observation of two solar mass neutron stars \citep[see also][]{Chamel2013,Alford:2013aca}. This conclusion was confirmed in \cite{Tews2018}, where the authors with just nuclear matter constraints obtained an increasing speed of sound with density.
In \cite{Tews2018} no peak was obtained around three times saturation density, as in studies where perturbative QCD constraints were also included \citep[][]{Annala2020,Altiparmak22,Gorda2022}. 
The behavior of speed of sound in our analysis is partly due to the underlying framework used to generate the neutron stars EOS which allows for different high density behaviors of the speed of sound, as discussed in \cite{Mueller:1996pm}. If the parameter $\zeta$ is large enough the high density behavior of the EOS changes from rising with the square of the density to increasing with a smaller power. 
The interesting result of our analysis is that observations favor large values of $\zeta$, and, therefore a speed of sound that saturates around $\sqrt{0.4}c$ instead of increasing to values close to $c$ as in \cite{Tews2018}. Besides, in a recent work it was shown  that EOS with large values of  $\zeta$ are compatible with the pQCD EOS \citep{Malik:2023mnx}. Our analysis of current observations requires a speed of sound larger than the conformal limit $c/\sqrt{3}$ in the interior of neutron stars, confirming the conclusions of \citet{Bedaque2015} and \citet{Tews2018}, but not an increasing function of density as in \citet{Tews2018}, nor with a sharp peak as in \citet{Gorda2022}.

\begin{figure}
	\centering
	\includegraphics[scale=0.38]{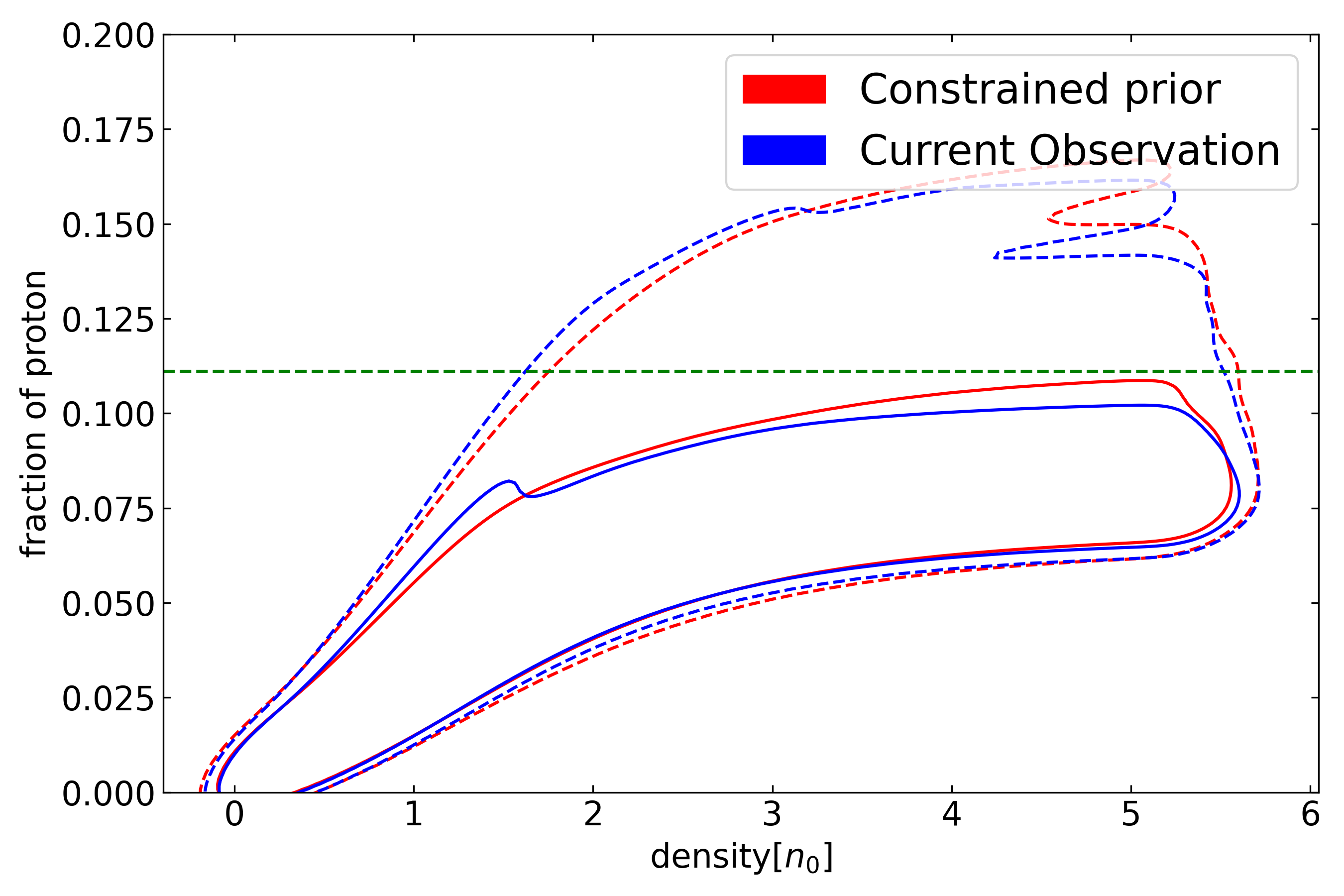}
	\caption{The posterior distribution of the proton fraction in neutron stars derived from current observations (blue solid line - 68\% credible interval, blue dashed line - 84\% credible interval), compared to the constrained prior distribution (same levels) in red. The green horizontal line defines the nucleonic direct Urca onset when muons are not considered, $y_p=1/9$. Note that the prominences at the boundaries of the contour levels are sampling artefacts.}. 
	
	\label{proton}
\end{figure} 

Performing inference using a real physics model framework gives some unique insights. One of these is the possible range of the total fraction of different particles in neutron stars. The proton fraction is of particular interest because it is directly related to the neutron star cooling process. As we are assuming beta equilibrium and electric neutrality, the proton fraction is equal to the sum of the electron and muon fractions. In Figure \ref{proton},  the posterior  (blue lines) and constrained prior contours (red lines) of the proton fraction are plotted as a function of the baryonic density. The horizontal line at $y_p=1/9$ indicates the onset of the nucleonic direct Urca processes when muons are not included. Including muons the nucleonic direct Urca sets in for $0.11<y_p \lesssim 0.14$.  We conclude that most of the models do not predict nucleonic direct Urca. However, after incorporating the constraints from current observations, the posterior distribution does not differ much from the constrained prior, suggesting that current astrophysical  observations do not have much constraining power on this quantity. The conclusion is consistent with the one drawn earlier regarding the ability to constrain the couplings associated with the isovector channel, $g_\rho$ and $\Lambda_{\omega}$. The distributions indicate that the posterior predicts a lower proton fraction than the constrained prior distribution when considering the 68\% credible level, which disfavors the existence of the direct Urca process. Notice, however, that direct Urca processes may still occur if hyperons set in \citep{Negreiros:2018cho,Providencia:2018ywl,Fortin:2020qin}.  In \citet{Beznogov:2015ewa},  it is predicted that according to observations,  direct Urca should open in stars with mass 1.6 to 1.8 \msol. A direct Urca constraint could easily be included in the analysis and will be considered in a future study.

\section{Conclusions}
\label{sec:conclusions}
In the present study, we have considered a microscopic nuclear model based on a field theoretical approach to span the whole meaningful neutron star M-R space. The model is based on a relativistic formulation and therefore has causality built-in. The parameters of the model have then been constrained by current neutron star observations, with only minimal guidance with respect to some nuclear matter properties: in particular, saturation density, binding energy at saturation and incompressibility.  This approach is in contrast to several commonly-used and more agnostic descriptions of the neutron star EOS, which do not contain information on the nuclear matter composition.

The model has been first constrained by current observations (radio, X-ray and gravitational waves). In a second step, we have studied how effectively future X-ray observations could constrain the model parameters.  We find that current observations mainly constrain the isoscalar channel of the EOS, i.e. the symmetric nuclear matter EOS.  It is interesting to note that NICER’s radius measurement for the high mass pulsar PSR J0740+6620 has a visible effect on constraining parameters. This leads us to conclude that it is important to have simultaneous information from both low and high mass stars to more efficiently obtain information on the EOS.

The isovector channel on the other hand - in particular, the parameters that define the density dependence of the symmetry energy - turned out to be less sensitive to the current observations and to our `Future' observational scenario (defined by M-R measurements at the $\sim 5$ \% level).  The reason could be simply a weakly constraining set of observations, since improved precision as simulated by our 'Future-X' scenario (M-R measurements at the $\sim 2$ \% level) seems to put some constraints, or it could have a more fundamental nature, such as the fact that the observation of mass and radius does not give information on the composition of matter.

We have also shown that future M-R observations with low uncertainty, of the type that we expect to be able to achieve using future large area X-ray timing telescopes, should allow us to make measurements of the EOS using only pulse profile modelling. In the scenario that we considered, measurements at the $\sim 2$ \% level were able to put strong constraints on most model parameters, in particular, on the parameters that determine the symmetric nuclear matter EOS behavior, and demonstrated robust recovery of the injected EOS model parameters.  Although we considered only two test cases in this study, this is very encouraging.  
Again, it will be important to have both high and low mass stars in the data set.  Importantly, with the capabilities that we anticipate for future instruments, we will be able to cross-check EOS inference derived solely from pulse profile modelling against that derived solely from GW measurements, thus allowing us to test for modelling and systematic errors in both techniques.  

Current observations also allow us to place constraints on the behaviour of both the speed of sound and the proton fraction. 
 It was found that the speed of sound squared saturates around $0.4c^2$. This is an interesting result because the model allows a speed of sound that tends at sufficiently large densities to a value between the conformal limit $c/\sqrt{3}$ and the speed of light $c$, depending on the magnitude of the coupling $\zeta$ of the quartic isoscalar–vector self-interaction term.  Observations seem to prefer larger values of $\zeta$ and, consequently, smaller speeds of sound and EOS that are compatible with the pQCD EOS. Finally,
 it has been shown that the proton fraction disfavors the nucleonic direct Urca, but it is important to stress that the isovector parameters are not very sensitive to the current observations used to extract proton fraction, and, therefore, this result is not strongly binding. 

\section*{Acknowledgements}

C.H. acknowledges support from an Arts \& Sciences Fellowship at Washington University in St. Louis, and would like to thank Yves Kini, Tuomo Salmi, Nathan Rutherford, Alex Chen, Serena Vinciguerra, Daniela Huppenkothen, Mark Alford, Alexander Haber, Liam Brodie and the anonymous referee for helpful discussions and comments. G.R. is grateful for financial support from the Nederlandse Organisatie voor Wetenschappelijk Onderzoek (NWO) through the Projectruimte and VIDI grants (PI S. Nissanke). A.L.W. acknowledges support from ERC Consolidator Grant (CoG) No.~865768 AEONS.  Computations were carried out on the HELIOS cluster on dedicated nodes funded via this grant. A.L.W. and G.R. would like to thank Jeannine de Kuijper, who did some preliminary work on this topic as part of her Bachelor and Masters degrees in Physics \& Astronomy at the University of Amsterdam/Vrije Universiteit Amsterdam.  L.T. acknowledges support from the  CEX2020-001058-M project (Unidad de Excelencia ``Mar\'{\i}a de Maeztu"), PID2019-110165GB-I00 project financed by the Spanish MCIN/ AEI/10.13039/501100011033/, the EU STRONG-2020 project under the program  H2020-INFRAIA-2018-1 grant agreement no. 824093, and from the Generalitat Valenciana under contract PROMETEO/2020/023.  C.P.  acknowledges support from FCT (Fundação para a Ciência e a Tecnologia, I.P, Portugal) under Projects  UIDP/\-04564/\-2020.  UIDB/\-04564/\-2020 and 2022.06460.PTDC.

\section*{Data Availability}

The posterior samples and scripts to make the plots in this paper are available in a Zenodo repository \citep{ChunHuang2023}

\bibliographystyle{mnras}
\bibliography{example} 

\begin{thebibliography}{}
\makeatletter
\relax
\def\mn@urlcharsother{\let\do\@makeother \do\$\do\&\do\#\do\^\do\_\do\%\do\~}
\def\mn@doi{\begingroup\mn@urlcharsother \@ifnextchar [ {\mn@doi@} {\mn@doi@[]}}
\def\mn@doi@[#1]#2{\def\@tempa{#1}\ifx\@tempa\@empty \href {http://dx.doi.org/#2} {doi:#2}\else \href {http://dx.doi.org/#2} {#1}\fi \endgroup}
\def\mn@eprint#1#2{\mn@eprint@#1:#2::\@nil}
\def\mn@eprint@arXiv#1{\href {http://arxiv.org/abs/#1} {{\tt arXiv:#1}}}
\def\mn@eprint@dblp#1{\href {http://dblp.uni-trier.de/rec/bibtex/#1.xml} {dblp:#1}}
\def\mn@eprint@#1:#2:#3:#4\@nil{\def\@tempa {#1}\def\@tempb {#2}\def\@tempc {#3}\ifx \@tempc \@empty \let \@tempc \@tempb \let \@tempb \@tempa \fi \ifx \@tempb \@empty \def\@tempb {arXiv}\fi \@ifundefined {mn@eprint@\@tempb}{\@tempb:\@tempc}{\expandafter \expandafter \csname mn@eprint@\@tempb\endcsname \expandafter{\@tempc}}}

\bibitem[\protect\citeauthoryear{Abbott et~al.}{Abbott et~al.}{2017}]{LIGOScientific:2017vwq}
Abbott B.~P.,  et~al., 2017, \mn@doi [\prl] {10.1103/PhysRevLett.119.161101}, 119, 161101

\bibitem[\protect\citeauthoryear{{Abbott} et~al.,}{{Abbott} et~al.}{2018}]{2018PhRvL.121p1101A}
{Abbott} B.~P.,  et~al., 2018, \mn@doi [\prl] {10.1103/PhysRevLett.121.161101}, \href {https://ui.adsabs.harvard.edu/abs/2018PhRvL.121p1101A} {121, 161101}

\bibitem[\protect\citeauthoryear{{Abbott} et~al.,}{{Abbott} et~al.}{2019}]{GW170817_TD2}
{Abbott} B.~P.,  et~al., 2019, \mn@doi [Phys. Rev. X.] {10.1103/PhysRevX.9.011001}, \href {https://ui.adsabs.harvard.edu/abs/2019PhRvX...9a1001A} {9, 011001}

\bibitem[\protect\citeauthoryear{Abbott et~al.}{Abbott et~al.}{2020}]{GW190425}
Abbott B.~P.,  et~al., 2020, \mn@doi [\apjl] {10.3847/2041-8213/ab75f5}, 892, L3

\bibitem[\protect\citeauthoryear{Adhikari et~al.}{Adhikari et~al.}{2021}]{PREX:2021umo}
Adhikari D.,  et~al., 2021, \mn@doi [Phys. Rev. Lett.] {10.1103/PhysRevLett.126.172502}, 126, 172502

\bibitem[\protect\citeauthoryear{Alam, Agrawal, Fortin, Pais, Provid\^encia, Raduta  \& Sulaksono}{Alam et~al.}{2016}]{Alam:2016cli}
Alam N.,  Agrawal B.~K.,  Fortin M.,  Pais H.,  Provid\^encia C.,  Raduta A.~R.,   Sulaksono A.,  2016, \mn@doi [Phys. Rev. C] {10.1103/PhysRevC.94.052801}, 94, 052801

\bibitem[\protect\citeauthoryear{Alford, Han  \& Prakash}{Alford et~al.}{2013}]{Alford:2013aca}
Alford M.~G.,  Han S.,   Prakash M.,  2013, \mn@doi [\prd] {10.1103/PhysRevD.88.083013}, 88, 083013

\bibitem[\protect\citeauthoryear{{Altiparmak}, {Ecker}  \& {Rezzolla}}{{Altiparmak} et~al.}{2022}]{Altiparmak22}
{Altiparmak} S.,  {Ecker} C.,   {Rezzolla} L.,  2022, \mn@doi [\apjl] {10.3847/2041-8213/ac9b2a}, \href {https://ui.adsabs.harvard.edu/abs/2022ApJ...939L..34A} {939, L34}

\bibitem[\protect\citeauthoryear{Annala, Gorda, Kurkela  \& Vuorinen}{Annala et~al.}{2018}]{Annala18}
Annala E.,  Gorda T.,  Kurkela A.,   Vuorinen A.,  2018, \mn@doi [\prl] {10.1103/PhysRevLett.120.172703}, 120, 172703

\bibitem[\protect\citeauthoryear{{Annala}, {Gorda}, {Kurkela}, {N{\"a}ttil{\"a}}  \& {Vuorinen}}{{Annala} et~al.}{2020}]{Annala2020}
{Annala} E.,  {Gorda} T.,  {Kurkela} A.,  {N{\"a}ttil{\"a}} J.,   {Vuorinen} A.,  2020, \mn@doi [Nature Physics] {10.1038/s41567-020-0914-9}, \href {https://ui.adsabs.harvard.edu/abs/2020NatPh..16..907A} {16, 907}

\bibitem[\protect\citeauthoryear{{Annala}, {Gorda}, {Katerini}, {Kurkela}, {N{\"a}ttil{\"a}}, {Paschalidis}  \& {Vuorinen}}{{Annala} et~al.}{2022}]{Annala2022}
{Annala} E.,  {Gorda} T.,  {Katerini} E.,  {Kurkela} A.,  {N{\"a}ttil{\"a}} J.,  {Paschalidis} V.,   {Vuorinen} A.,  2022, \mn@doi [Physical Review X] {10.1103/PhysRevX.12.011058}, \href {https://ui.adsabs.harvard.edu/abs/2022PhRvX..12a1058A} {12, 011058}

\bibitem[\protect\citeauthoryear{{Annala}, {Gorda}, {Hirvonen}, {Komoltsev}, {Kurkela}, {N{\"a}ttil{\"a}}  \& {Vuorinen}}{{Annala} et~al.}{2023}]{Annala23}
{Annala} E.,  {Gorda} T.,  {Hirvonen} J.,  {Komoltsev} O.,  {Kurkela} A.,  {N{\"a}ttil{\"a}} J.,   {Vuorinen} A.,  2023, \mn@doi [Nature Communications] {10.1038/s41467-023-44051-y}, \href {https://ui.adsabs.harvard.edu/abs/2023NatCo..14.8451A} {14, 8451}

\bibitem[\protect\citeauthoryear{Bao, Hu, Zhang  \& Shen}{Bao et~al.}{2014}]{Bao:2014lqa}
Bao S.~S.,  Hu J.~N.,  Zhang Z.~W.,   Shen H.,  2014, \mn@doi [\prc] {10.1103/PhysRevC.90.045802}, 90, 045802

\bibitem[\protect\citeauthoryear{Baym, Pethick  \& Sutherland}{Baym et~al.}{1971}]{Baym:1971pw}
Baym G.,  Pethick C.,   Sutherland P.,  1971, \mn@doi [\apj] {10.1086/151216}, 170, 299

\bibitem[\protect\citeauthoryear{{Baym}, {Hatsuda}, {Kojo}, {Powell}, {Song}  \& {Takatsuka}}{{Baym} et~al.}{2018}]{Baym2018}
{Baym} G.,  {Hatsuda} T.,  {Kojo} T.,  {Powell} P.~D.,  {Song} Y.,   {Takatsuka} T.,  2018, \mn@doi [Reports on Progress in Physics] {10.1088/1361-6633/aaae14}, \href {https://ui.adsabs.harvard.edu/abs/2018RPPh...81e6902B} {81, 056902}

\bibitem[\protect\citeauthoryear{{Bedaque} \& {Steiner}}{{Bedaque} \& {Steiner}}{2015}]{Bedaque2015}
{Bedaque} P.,  {Steiner} A.~W.,  2015, \mn@doi [\prl] {10.1103/PhysRevLett.114.031103}, \href {https://ui.adsabs.harvard.edu/abs/2015PhRvL.114c1103B} {114, 031103}

\bibitem[\protect\citeauthoryear{Beznogov \& Yakovlev}{Beznogov \& Yakovlev}{2015}]{Beznogov:2015ewa}
Beznogov M.~V.,  Yakovlev D.~G.,  2015, \mn@doi [\mnras] {10.1093/mnras/stv1293}, 452, 540

\bibitem[\protect\citeauthoryear{{Biswas}}{{Biswas}}{2022}]{Biswas2022}
{Biswas} B.,  2022, \mn@doi [\apj] {10.3847/1538-4357/ac447b}, \href {https://ui.adsabs.harvard.edu/abs/2022ApJ...926...75B} {926, 75}

\bibitem[\protect\citeauthoryear{Bodmer}{Bodmer}{1991}]{BODMER1991703}
Bodmer A.,  1991, \mn@doi [Nuclear Physics A] {https://doi.org/10.1016/0375-9474(91)90439-D}, 526, 703

\bibitem[\protect\citeauthoryear{Boguta \& Bodmer}{Boguta \& Bodmer}{1977}]{BOGUTA1977413}
Boguta J.,  Bodmer A.,  1977, \mn@doi [Nuclear Physics A] {https://doi.org/10.1016/0375-9474(77)90626-1}, 292, 413

\bibitem[\protect\citeauthoryear{Boguta \& Stocker}{Boguta \& Stocker}{1983}]{BOGUTA1983289}
Boguta J.,  Stocker H.,  1983, \mn@doi [Physics Letters B] {https://doi.org/10.1016/0370-2693(83)90446-X}, 120, 289

\bibitem[\protect\citeauthoryear{{Buchner}}{{Buchner}}{2016}]{2016S&C....26..383B}
{Buchner} J.,  2016, \mn@doi [Statistics and Computing] {10.1007/s11222-014-9512-y}, \href {https://ui.adsabs.harvard.edu/abs/2016S&C....26..383B} {26, 383}

\bibitem[\protect\citeauthoryear{{Buchner}}{{Buchner}}{2019}]{2019PASP..131j8005B}
{Buchner} J.,  2019, \mn@doi [\pasp] {10.1088/1538-3873/aae7fc}, \href {https://ui.adsabs.harvard.edu/abs/2019PASP..131j8005B} {131, 108005}

\bibitem[\protect\citeauthoryear{{Buchner}}{{Buchner}}{2021}]{2021JOSS....6.3001B}
{Buchner} J.,  2021, \mn@doi [The Journal of Open Source Software] {10.21105/joss.03001}, \href {https://ui.adsabs.harvard.edu/abs/2021JOSS....6.3001B} {6, 3001}

\bibitem[\protect\citeauthoryear{Burgio, Schulze, Vidana  \& Wei}{Burgio et~al.}{2021}]{Burgio:2021vgk}
Burgio G.~F.,  Schulze H.~J.,  Vidana I.,   Wei J.~B.,  2021, \mn@doi [Prog. Part. Nucl. Phys.] {10.1016/j.ppnp.2021.103879}, 120, 103879

\bibitem[\protect\citeauthoryear{Carriere, Horowitz  \& Piekarewicz}{Carriere et~al.}{2003}]{Carriere:2002bx}
Carriere J.,  Horowitz C.~J.,   Piekarewicz J.,  2003, \mn@doi [\apj] {10.1086/376515}, 593, 463

\bibitem[\protect\citeauthoryear{Cavagnoli, Menezes  \& Providencia}{Cavagnoli et~al.}{2011}]{Cavagnoli:2011ft}
Cavagnoli R.,  Menezes D.~P.,   Providencia C.,  2011, \mn@doi [\prc] {10.1103/PhysRevC.84.065810}, 84, 065810

\bibitem[\protect\citeauthoryear{{Chamel}, {Fantina}, {Pearson}  \& {Goriely}}{{Chamel} et~al.}{2013}]{Chamel2013}
{Chamel} N.,  {Fantina} A.~F.,  {Pearson} J.~M.,   {Goriely} S.,  2013, \mn@doi [\aap] {10.1051/0004-6361/201220986}, \href {https://ui.adsabs.harvard.edu/abs/2013A&A...553A..22C} {553, A22}

\bibitem[\protect\citeauthoryear{{Chen} \& {Piekarewicz}}{{Chen} \& {Piekarewicz}}{2014}]{Chen_2014}
{Chen} W.-C.,  {Piekarewicz} J.,  2014, \mn@doi [\prc] {10.1103/PhysRevC.90.044305}, \href {https://ui.adsabs.harvard.edu/abs/2014PhRvC..90d4305C} {90, 044305}

\bibitem[\protect\citeauthoryear{{Cromartie} et~al.,}{{Cromartie} et~al.}{2020}]{Cromartie2020}
{Cromartie} H.~T.,  et~al., 2020, \mn@doi [Nature Astronomy] {10.1038/s41550-019-0880-2}, \href {https://ui.adsabs.harvard.edu/abs/2020NatAs...4...72C} {4, 72}

\bibitem[\protect\citeauthoryear{{Demorest}, {Pennucci}, {Ransom}, {Roberts}  \& {Hessels}}{{Demorest} et~al.}{2010}]{Demorest2010}
{Demorest} P.~B.,  {Pennucci} T.,  {Ransom} S.~M.,  {Roberts} M.~S.~E.,   {Hessels} J.~W.~T.,  2010, \mn@doi [\nat] {10.1038/nature09466}, \href {https://ui.adsabs.harvard.edu/abs/2010Natur.467.1081D} {467, 1081}

\bibitem[\protect\citeauthoryear{Dutra et~al.,}{Dutra et~al.}{2014}]{Dutra:2014qga}
Dutra M.,  et~al., 2014, \mn@doi [\prc] {10.1103/PhysRevC.90.055203}, 90, 055203

\bibitem[\protect\citeauthoryear{Essick, Tews, Landry  \& Schwenk}{Essick et~al.}{2021}]{Essick2021}
Essick R.,  Tews I.,  Landry P.,   Schwenk A.,  2021, \mn@doi [\prl] {10.1103/PhysRevLett.127.192701}, 127, 192701

\bibitem[\protect\citeauthoryear{Fattoyev \& Piekarewicz}{Fattoyev \& Piekarewicz}{2010a}]{Fattoyev:2010rx}
Fattoyev F.~J.,  Piekarewicz J.,  2010a, \mn@doi [Phys. Rev. C] {10.1103/PhysRevC.82.025805}, 82, 025805

\bibitem[\protect\citeauthoryear{Fattoyev \& Piekarewicz}{Fattoyev \& Piekarewicz}{2010b}]{Fattoyev:2010tb}
Fattoyev F.~J.,  Piekarewicz J.,  2010b, \mn@doi [Phys. Rev. C] {10.1103/PhysRevC.82.025810}, 82, 025810

\bibitem[\protect\citeauthoryear{Fattoyev, Horowitz, Piekarewicz  \& Shen}{Fattoyev et~al.}{2010}]{Fattoyev:2010mx}
Fattoyev F.~J.,  Horowitz C.~J.,  Piekarewicz J.,   Shen G.,  2010, \mn@doi [\prc] {10.1103/PhysRevC.82.055803}, 82, 055803

\bibitem[\protect\citeauthoryear{Fattoyev, Horowitz, Piekarewicz  \& Reed}{Fattoyev et~al.}{2020}]{Fattoyev:2020cws}
Fattoyev F.~J.,  Horowitz C.~J.,  Piekarewicz J.,   Reed B.,  2020, \mn@doi [\prc] {10.1103/PhysRevC.102.065805}, 102, 065805

\bibitem[\protect\citeauthoryear{{Fonseca} et~al.,}{{Fonseca} et~al.}{2021}]{Fonseca_2021}
{Fonseca} E.,  et~al., 2021, \mn@doi [\apjl] {10.3847/2041-8213/ac03b8}, \href {https://ui.adsabs.harvard.edu/abs/2021ApJ...915L..12F} {915, L12}

\bibitem[\protect\citeauthoryear{Fortin, Providencia, Raduta, Gulminelli, Zdunik, Haensel  \& Bejger}{Fortin et~al.}{2016}]{Fortin:2016hny}
Fortin M.,  Providencia C.,  Raduta A.~R.,  Gulminelli F.,  Zdunik J.~L.,  Haensel P.,   Bejger M.,  2016, \mn@doi [Phys. Rev. C] {10.1103/PhysRevC.94.035804}, 94, 035804

\bibitem[\protect\citeauthoryear{Fortin, Raduta, Avancini  \& Provid\^encia}{Fortin et~al.}{2020}]{Fortin:2020qin}
Fortin M.,  Raduta A.~R.,  Avancini S.,   Provid\^encia C.,  2020, \mn@doi [\prd] {10.1103/PhysRevD.101.034017}, 101, 034017

\bibitem[\protect\citeauthoryear{{Gendreau} et~al.,}{{Gendreau} et~al.}{2016}]{Gendreau2016}
{Gendreau} K.~C.,  et~al., 2016, in {den Herder} J.-W.~A.,  {Takahashi} T.,   {Bautz} M.,  eds,  Society of Photo-Optical Instrumentation Engineers (SPIE) Conference Series Vol. 9905, Space Telescopes and Instrumentation 2016: Ultraviolet to Gamma Ray. p. 99051H, \mn@doi{10.1117/12.2231304}

\bibitem[\protect\citeauthoryear{{Ghosh}, {Pradhan}, {Chatterjee}  \& {Schaffner-Bielich}}{{Ghosh} et~al.}{2022a}]{Ghosh_2022_1}
{Ghosh} S.,  {Pradhan} B.~K.,  {Chatterjee} D.,   {Schaffner-Bielich} J.,  2022a, \mn@doi [Frontiers in Astronomy and Space Sciences] {10.3389/fspas.2022.864294}, \href {https://ui.adsabs.harvard.edu/abs/2022FrASS...964294G} {9, 864294}

\bibitem[\protect\citeauthoryear{{Ghosh}, {Chatterjee}  \& {Schaffner-Bielich}}{{Ghosh} et~al.}{2022b}]{Ghosh_2022}
{Ghosh} S.,  {Chatterjee} D.,   {Schaffner-Bielich} J.,  2022b, \mn@doi [European Physical Journal A] {10.1140/epja/s10050-022-00679-w}, \href {https://ui.adsabs.harvard.edu/abs/2022EPJA...58...37G} {58, 37}

\bibitem[\protect\citeauthoryear{{Glendenning}}{{Glendenning}}{1996}]{Glendenning:1997wn}
{Glendenning} N.,  1996, {Compact Stars. Nuclear Physics, Particle Physics and General Relativity.}.
Springer-Verlag New York

\bibitem[\protect\citeauthoryear{{Gorda}, {Komoltsev}  \& {Kurkela}}{{Gorda} et~al.}{2023}]{Gorda2022}
{Gorda} T.,  {Komoltsev} O.,   {Kurkela} A.,  2023, \mn@doi [\apj] {10.3847/1538-4357/acce3a}, \href {https://ui.adsabs.harvard.edu/abs/2023ApJ...950..107G} {950, 107}

\bibitem[\protect\citeauthoryear{{Greif}, {Raaijmakers}, {Hebeler}, {Schwenk}  \& {Watts}}{{Greif} et~al.}{2019}]{Greif_2019}
{Greif} S.~K.,  {Raaijmakers} G.,  {Hebeler} K.,  {Schwenk} A.,   {Watts} A.~L.,  2019, \mn@doi [\mnras] {10.1093/mnras/stz654}, \href {https://ui.adsabs.harvard.edu/abs/2019MNRAS.485.5363G} {485, 5363}

\bibitem[\protect\citeauthoryear{{Hebeler}}{{Hebeler}}{2021}]{Hebeler2021}
{Hebeler} K.,  2021, \mn@doi [\physrep] {10.1016/j.physrep.2020.08.009}, \href {https://ui.adsabs.harvard.edu/abs/2021PhR...890....1H} {890, 1}

\bibitem[\protect\citeauthoryear{Hornick, Tolos, Zacchi, Christian  \& Schaffner-Bielich}{Hornick et~al.}{2018}]{Hornick:2018kfi}
Hornick N.,  Tolos L.,  Zacchi A.,  Christian J.-E.,   Schaffner-Bielich J.,  2018, \mn@doi [\prc] {10.1103/PhysRevC.98.065804}, 98, 065804

\bibitem[\protect\citeauthoryear{Horowitz \& Piekarewicz}{Horowitz \& Piekarewicz}{2001a}]{Horowitz:2001ya}
Horowitz C.~J.,  Piekarewicz J.,  2001a, \mn@doi [Phys. Rev. C] {10.1103/PhysRevC.64.062802}, 64, 062802

\bibitem[\protect\citeauthoryear{Horowitz \& Piekarewicz}{Horowitz \& Piekarewicz}{2001b}]{Horowitz:2000xj}
Horowitz C.~J.,  Piekarewicz J.,  2001b, \mn@doi [\prl] {10.1103/PhysRevLett.86.5647}, 86, 5647

\bibitem[\protect\citeauthoryear{{Huang}, {Raaijmakers}, {Watts}, {Tolos}  \& {Provindencia}}{{Huang} et~al.}{2023}]{ChunHuang2023}
{Huang} C.,  {Raaijmakers} G.,  {Watts} A.~L.,  {Tolos} L.,   {Provindencia} C.,  2023, Constraining fundamental nuclear physics parameters using neutron star mass-radius measurements I: Nucleonic models, \mn@doi{10.5281/zenodo.7551909}

\bibitem[\protect\citeauthoryear{Huth, Wellenhofer  \& Schwenk}{Huth et~al.}{2021}]{Huth:2020ozf}
Huth S.,  Wellenhofer C.,   Schwenk A.,  2021, \mn@doi [\prc] {10.1103/PhysRevC.103.025803}, 103, 025803

\bibitem[\protect\citeauthoryear{Kass \& Raftery}{Kass \& Raftery}{1995}]{Kass}
Kass R.~E.,  Raftery A.~E.,  1995, Journal of the American Statistical Association, 90, 773

\bibitem[\protect\citeauthoryear{Kurkela, Fraga, Schaffner-Bielich  \& Vuorinen}{Kurkela et~al.}{2014}]{Kurkela:2014vha}
Kurkela A.,  Fraga E.~S.,  Schaffner-Bielich J.,   Vuorinen A.,  2014, \mn@doi [\apj] {10.1088/0004-637X/789/2/127}, 789, 127

\bibitem[\protect\citeauthoryear{Landry, Essick  \& Chatziioannou}{Landry et~al.}{2020}]{Landry:2020vaw}
Landry P.,  Essick R.,   Chatziioannou K.,  2020, \mn@doi [\prd] {10.1103/PhysRevD.101.123007}, 101, 123007

\bibitem[\protect\citeauthoryear{{Lattimer}}{{Lattimer}}{2012}]{Lattimer12ARNPS}
{Lattimer} J.~M.,  2012, \mn@doi [Annual Review of Nuclear and Particle Science] {10.1146/annurev-nucl-102711-095018}, \href {http://adsabs.harvard.edu/abs/2012ARNPS..62..485L} {62, 485}

\bibitem[\protect\citeauthoryear{Lattimer \& Lim}{Lattimer \& Lim}{2013}]{Lattimer:2012xj}
Lattimer J.~M.,  Lim Y.,  2013, \mn@doi [\apj] {10.1088/0004-637X/771/1/51}, 771, 51

\bibitem[\protect\citeauthoryear{{Legred}, {Chatziioannou}, {Essick}, {Han}  \& {Landry}}{{Legred} et~al.}{2021}]{Legred21}
{Legred} I.,  {Chatziioannou} K.,  {Essick} R.,  {Han} S.,   {Landry} P.,  2021, \mn@doi [\prd] {10.1103/PhysRevD.104.063003}, \href {https://ui.adsabs.harvard.edu/abs/2021PhRvD.104f3003L} {104, 063003}

\bibitem[\protect\citeauthoryear{{Legred}, {Chatziioannou}, {Essick}  \& {Landry}}{{Legred} et~al.}{2022}]{Legred2022}
{Legred} I.,  {Chatziioannou} K.,  {Essick} R.,   {Landry} P.,  2022, \mn@doi [\prd] {10.1103/PhysRevD.105.043016}, \href {https://ui.adsabs.harvard.edu/abs/2022PhRvD.105d3016L} {105, 043016}

\bibitem[\protect\citeauthoryear{{Li}, {Sedrakian}  \& {Alford}}{{Li} et~al.}{2021}]{JieLiJ21}
{Li} J.~J.,  {Sedrakian} A.,   {Alford} M.,  2021, \mn@doi [\prd] {10.1103/PhysRevD.104.L121302}, \href {https://ui.adsabs.harvard.edu/abs/2021PhRvD.104l1302L} {104, L121302}

\bibitem[\protect\citeauthoryear{{Lindblom}}{{Lindblom}}{2010}]{Lindblom2010}
{Lindblom} L.,  2010, \mn@doi [\prd] {10.1103/PhysRevD.82.103011}, \href {https://ui.adsabs.harvard.edu/abs/2010PhRvD..82j3011L} {82, 103011}

\bibitem[\protect\citeauthoryear{{Lindblom}}{{Lindblom}}{2022}]{Lindblom2022}
{Lindblom} L.,  2022, \mn@doi [\prd] {10.1103/PhysRevD.105.063031}, \href {https://ui.adsabs.harvard.edu/abs/2022PhRvD.105f3031L} {105, 063031}

\bibitem[\protect\citeauthoryear{{Lo}, {Miller}, {Bhattacharyya}  \& {Lamb}}{{Lo} et~al.}{2013}]{Lo2013}
{Lo} K.~H.,  {Miller} M.~C.,  {Bhattacharyya} S.,   {Lamb} F.~K.,  2013, \mn@doi [\apj] {10.1088/0004-637X/776/1/19}, \href {https://ui.adsabs.harvard.edu/abs/2013ApJ...776...19L} {776, 19}

\bibitem[\protect\citeauthoryear{Malik, Alam, Fortin, Provid\^encia, Agrawal, Jha, Kumar  \& Patra}{Malik et~al.}{2018}]{Malik:2018zcf}
Malik T.,  Alam N.,  Fortin M.,  Provid\^encia C.,  Agrawal B.~K.,  Jha T.~K.,  Kumar B.,   Patra S.~K.,  2018, \mn@doi [Phys. Rev. C] {10.1103/PhysRevC.98.035804}, 98, 035804

\bibitem[\protect\citeauthoryear{Malik, Ferreira, Agrawal  \& Provid\^encia}{Malik et~al.}{2022}]{Malik:2022zol}
Malik T.,  Ferreira M.,  Agrawal B.~K.,   Provid\^encia C.,  2022, \mn@doi [\apj] {10.3847/1538-4357/ac5d3c}, 930, 17

\bibitem[\protect\citeauthoryear{{Malik}, {Ferreira}, {Albino}  \& {Provid{\^e}ncia}}{{Malik} et~al.}{2023}]{Malik:2023mnx}
{Malik} T.,  {Ferreira} M.,  {Albino} M.~B.,   {Provid{\^e}ncia} C.,  2023, \mn@doi [\prd] {10.1103/PhysRevD.107.103018}, \href {https://ui.adsabs.harvard.edu/abs/2023PhRvD.107j3018M} {107, 103018}

\bibitem[\protect\citeauthoryear{Margueron, Hoffmann~Casali  \& Gulminelli}{Margueron et~al.}{2018}]{Margueron:2017eqc}
Margueron J.,  Hoffmann~Casali R.,   Gulminelli F.,  2018, \mn@doi [\prc] {10.1103/PhysRevC.97.025805}, 97, 025805

\bibitem[\protect\citeauthoryear{{Miller} et~al.,}{{Miller} et~al.}{2019}]{Miller2019}
{Miller} M.~C.,  et~al., 2019, \mn@doi [\apjl] {10.3847/2041-8213/ab50c5}, \href {https://ui.adsabs.harvard.edu/abs/2019ApJ...887L..24M} {887, L24}

\bibitem[\protect\citeauthoryear{{Miller} et~al.,}{{Miller} et~al.}{2021}]{Miller2021}
{Miller} M.~C.,  et~al., 2021, \mn@doi [\apjl] {10.3847/2041-8213/ac089b}, \href {https://ui.adsabs.harvard.edu/abs/2021ApJ...918L..28M} {918, L28}

\bibitem[\protect\citeauthoryear{Mondal \& Gulminelli}{Mondal \& Gulminelli}{2023}]{Mondal:2022cva}
Mondal C.,  Gulminelli F.,  2023, \mn@doi [Phys. Rev. C] {10.1103/PhysRevC.107.015801}, 107, 015801

\bibitem[\protect\citeauthoryear{Most, Weih, Rezzolla  \& Schaffner-Bielich}{Most et~al.}{2018}]{Most2018}
Most E.~R.,  Weih L.~R.,  Rezzolla L.,   Schaffner-Bielich J.,  2018, \mn@doi [\prl] {10.1103/PhysRevLett.120.261103}, 120, 261103

\bibitem[\protect\citeauthoryear{{M{\"u}ller} \& {Serot}}{{M{\"u}ller} \& {Serot}}{1996}]{Mueller:1996pm}
{M{\"u}ller} H.,  {Serot} B.~D.,  1996, \mn@doi [\nphysa] {10.1016/0375-9474(96)00187-X}, \href {https://ui.adsabs.harvard.edu/abs/1996NuPhA.606..508M} {606, 508}

\bibitem[\protect\citeauthoryear{Negreiros, Tolos, Centelles, Ramos  \& Dexheimer}{Negreiros et~al.}{2018}]{Negreiros:2018cho}
Negreiros R.,  Tolos L.,  Centelles M.,  Ramos A.,   Dexheimer V.,  2018, \mn@doi [Astrophys. J.] {10.3847/1538-4357/aad049}, 863, 104

\bibitem[\protect\citeauthoryear{{Oertel}, {Hempel}, {Kl{\"a}hn}  \& {Typel}}{{Oertel} et~al.}{2017}]{OE17}
{Oertel} M.,  {Hempel} M.,  {Kl{\"a}hn} T.,   {Typel} S.,  2017, \mn@doi [Reviews of Modern Physics] {10.1103/RevModPhys.89.015007}, \href {https://ui.adsabs.harvard.edu/abs/2017RvMP...89a5007O} {89, 015007}

\bibitem[\protect\citeauthoryear{{Oppenheimer} \& {Volkoff}}{{Oppenheimer} \& {Volkoff}}{1939}]{Oppenheimer39}
{Oppenheimer} J.~R.,  {Volkoff} G.~M.,  1939, \mn@doi [Physical Review] {10.1103/PhysRev.55.374}, \href {https://ui.adsabs.harvard.edu/abs/1939PhRv...55..374O} {55, 374}

\bibitem[\protect\citeauthoryear{{{\"O}zel}, {Psaltis}, {Ransom}, {Demorest}  \& {Alford}}{{{\"O}zel} et~al.}{2010}]{Ozel2010}
{{\"O}zel} F.,  {Psaltis} D.,  {Ransom} S.,  {Demorest} P.,   {Alford} M.,  2010, \mn@doi [\apjl] {10.1088/2041-8205/724/2/L199}, \href {https://ui.adsabs.harvard.edu/abs/2010ApJ...724L.199O} {724, L199}

\bibitem[\protect\citeauthoryear{Pais \& Provid\^encia}{Pais \& Provid\^encia}{2016}]{Pais:2016xiu}
Pais H.,  Provid\^encia C.,  2016, \mn@doi [\prc] {10.1103/PhysRevC.94.015808}, 94, 015808

\bibitem[\protect\citeauthoryear{{Pang}, {Tews}, {Coughlin}, {Bulla}, {Van Den Broeck}  \& {Dietrich}}{{Pang} et~al.}{2021}]{Pang21}
{Pang} P. T.~H.,  {Tews} I.,  {Coughlin} M.~W.,  {Bulla} M.,  {Van Den Broeck} C.,   {Dietrich} T.,  2021, \mn@doi [\apj] {10.3847/1538-4357/ac19ab}, \href {https://ui.adsabs.harvard.edu/abs/2021ApJ...922...14P} {922, 14}

\bibitem[\protect\citeauthoryear{Piekarewicz, Fattoyev  \& Horowitz}{Piekarewicz et~al.}{2014}]{Piekarewicz:2014lba}
Piekarewicz J.,  Fattoyev F.~J.,   Horowitz C.~J.,  2014, \mn@doi [\prc] {10.1103/PhysRevC.90.015803}, 90, 015803

\bibitem[\protect\citeauthoryear{Providencia \& Rabhi}{Providencia \& Rabhi}{2013}]{Providencia:2012rx}
Providencia C.,  Rabhi A.,  2013, \mn@doi [\prc] {10.1103/PhysRevC.87.055801}, 87, 055801

\bibitem[\protect\citeauthoryear{{Provid{\^e}ncia}, {Fortin}, {Pais}  \& {Rabhi}}{{Provid{\^e}ncia} et~al.}{2019}]{Providencia:2018ywl}
{Provid{\^e}ncia} C.,  {Fortin} M.,  {Pais} H.,   {Rabhi} A.,  2019, \mn@doi [Frontiers in Astronomy and Space Sciences] {10.3389/fspas.2019.00013}, \href {https://ui.adsabs.harvard.edu/abs/2019FrASS...6...13P} {6, 13}

\bibitem[\protect\citeauthoryear{{Psaltis}, {{\"O}zel}  \& {Chakrabarty}}{{Psaltis} et~al.}{2014}]{Psaltis2014}
{Psaltis} D.,  {{\"O}zel} F.,   {Chakrabarty} D.,  2014, \mn@doi [\apj] {10.1088/0004-637X/787/2/136}, \href {https://ui.adsabs.harvard.edu/abs/2014ApJ...787..136P} {787, 136}

\bibitem[\protect\citeauthoryear{{Raaijmakers} et~al.,}{{Raaijmakers} et~al.}{2019}]{Raaijmakers_2019}
{Raaijmakers} G.,  et~al., 2019, \mn@doi [\apjl] {10.3847/2041-8213/ab451a}, \href {https://ui.adsabs.harvard.edu/abs/2019ApJ...887L..22R} {887, L22}

\bibitem[\protect\citeauthoryear{{Raaijmakers} et~al.,}{{Raaijmakers} et~al.}{2020}]{Raaijmakers_2020}
{Raaijmakers} G.,  et~al., 2020, \mn@doi [\apjl] {10.3847/2041-8213/ab822f}, \href {https://ui.adsabs.harvard.edu/abs/2020ApJ...893L..21R} {893, L21}

\bibitem[\protect\citeauthoryear{Raaijmakers et~al.,}{Raaijmakers et~al.}{2021}]{Raaijmakers_2021}
Raaijmakers G.,  et~al., 2021, \mn@doi [\apjl] {10.3847/2041-8213/ac089a}, 918, L29

\bibitem[\protect\citeauthoryear{{Ray} et~al.,}{{Ray} et~al.}{2019}]{strobex}
{Ray} P.~S.,  et~al., 2019, arXiv e-prints, \href {https://ui.adsabs.harvard.edu/abs/2019arXiv190303035R} {p. arXiv:1903.03035}

\bibitem[\protect\citeauthoryear{{Read}, {Lackey}, {Owen}  \& {Friedman}}{{Read} et~al.}{2009}]{Read2009}
{Read} J.~S.,  {Lackey} B.~D.,  {Owen} B.~J.,   {Friedman} J.~L.,  2009, \mn@doi [\prd] {10.1103/PhysRevD.79.124032}, \href {https://ui.adsabs.harvard.edu/abs/2009PhRvD..79l4032R} {79, 124032}

\bibitem[\protect\citeauthoryear{{Reardon} et~al.,}{{Reardon} et~al.}{2016}]{Reardon2016}
{Reardon} D.~J.,  et~al., 2016, \mn@doi [\mnras] {10.1093/mnras/stv2395}, \href {https://ui.adsabs.harvard.edu/abs/2016MNRAS.455.1751R} {455, 1751}

\bibitem[\protect\citeauthoryear{Reed, Fattoyev, Horowitz  \& Piekarewicz}{Reed et~al.}{2021}]{Reed2021}
Reed B.~T.,  Fattoyev F.~J.,  Horowitz C.~J.,   Piekarewicz J.,  2021, \mn@doi [Phys. Rev. Lett.] {10.1103/PhysRevLett.126.172503}, 126, 172503

\bibitem[\protect\citeauthoryear{Reinhard, Roca-Maza  \& Nazarewicz}{Reinhard et~al.}{2021}]{Reinhard:2021utv}
Reinhard P.-G.,  Roca-Maza X.,   Nazarewicz W.,  2021, \mn@doi [Phys. Rev. Lett.] {10.1103/PhysRevLett.127.232501}, 127, 232501

\bibitem[\protect\citeauthoryear{Reinhard, Roca-Maza  \& Nazarewicz}{Reinhard et~al.}{2022}]{Reinhard:2022inh}
Reinhard P.-G.,  Roca-Maza X.,   Nazarewicz W.,  2022, \mn@doi [Phys. Rev. Lett.] {10.1103/PhysRevLett.129.232501}, 129, 232501

\bibitem[\protect\citeauthoryear{{Riley} et~al.,}{{Riley} et~al.}{2019}]{Riley2019}
{Riley} T.~E.,  et~al., 2019, \mn@doi [\apjl] {10.3847/2041-8213/ab481c}, \href {https://ui.adsabs.harvard.edu/abs/2019ApJ...887L..21R} {887, L21}

\bibitem[\protect\citeauthoryear{{Riley} et~al.,}{{Riley} et~al.}{2021}]{Riley2021}
{Riley} T.~E.,  et~al., 2021, \mn@doi [\apjl] {10.3847/2041-8213/ac0a81}, \href {https://ui.adsabs.harvard.edu/abs/2021ApJ...918L..27R} {918, L27}

\bibitem[\protect\citeauthoryear{{Rutherford}, {Raaijmakers}, {Prescod-Weinstein}  \& {Watts}}{{Rutherford} et~al.}{2023}]{Rutherford22}
{Rutherford} N.,  {Raaijmakers} G.,  {Prescod-Weinstein} C.,   {Watts} A.,  2023, \mn@doi [\prd] {10.1103/PhysRevD.107.103051}, \href {https://ui.adsabs.harvard.edu/abs/2023PhRvD.107j3051R} {107, 103051}

\bibitem[\protect\citeauthoryear{{Salmi} et~al.,}{{Salmi} et~al.}{2022}]{Salmi22}
{Salmi} T.,  et~al., 2022, \mn@doi [\apj] {10.3847/1538-4357/ac983d}, \href {https://ui.adsabs.harvard.edu/abs/2022ApJ...941..150S} {941, 150}

\bibitem[\protect\citeauthoryear{Serot \& Walecka}{Serot \& Walecka}{1986}]{Serot:1984ey}
Serot B.~D.,  Walecka J.~D.,  1986, Adv. Nucl. Phys., 16, 1

\bibitem[\protect\citeauthoryear{Shen, Ji, Hu  \& Sumiyoshi}{Shen et~al.}{2020}]{Shen:2020sec}
Shen H.,  Ji F.,  Hu J.,   Sumiyoshi K.,  2020, \mn@doi [\apj] {10.3847/1538-4357/ab72fd}, 891, 148

\bibitem[\protect\citeauthoryear{Stone, Stone  \& Moszkowski}{Stone et~al.}{2014}]{Stone:2014wza}
Stone J.~R.,  Stone N.~J.,   Moszkowski S.~A.,  2014, \mn@doi [\prc] {10.1103/PhysRevC.89.044316}, 89, 044316

\bibitem[\protect\citeauthoryear{{Sun}, {Miao}, {Sun}  \& {Li}}{{Sun} et~al.}{2023}]{Sun2022}
{Sun} X.,  {Miao} Z.,  {Sun} B.,   {Li} A.,  2023, \mn@doi [\apj] {10.3847/1538-4357/ac9d9a}, \href {https://ui.adsabs.harvard.edu/abs/2023ApJ...942...55S} {942, 55}

\bibitem[\protect\citeauthoryear{{Tang}, {Jiang}, {Han}, {Fan}  \& {Wei}}{{Tang} et~al.}{2021}]{TangSP21}
{Tang} S.-P.,  {Jiang} J.-L.,  {Han} M.-Z.,  {Fan} Y.-Z.,   {Wei} D.-M.,  2021, \mn@doi [\prd] {10.1103/PhysRevD.104.063032}, \href {https://ui.adsabs.harvard.edu/abs/2021PhRvD.104f3032T} {104, 063032}

\bibitem[\protect\citeauthoryear{{Tews}, {Carlson}, {Gandolfi}  \& {Reddy}}{{Tews} et~al.}{2018}]{Tews2018}
{Tews} I.,  {Carlson} J.,  {Gandolfi} S.,   {Reddy} S.,  2018, \mn@doi [\apj] {10.3847/1538-4357/aac267}, \href {https://ui.adsabs.harvard.edu/abs/2018ApJ...860..149T} {860, 149}

\bibitem[\protect\citeauthoryear{Todd-Rutel \& Piekarewicz}{Todd-Rutel \& Piekarewicz}{2005}]{Todd-Rutel:2005yzo}
Todd-Rutel B.~G.,  Piekarewicz J.,  2005, \mn@doi [\prl] {10.1103/PhysRevLett.95.122501}, 95, 122501

\bibitem[\protect\citeauthoryear{{Tolman}}{{Tolman}}{1939}]{Tolman39}
{Tolman} R.~C.,  1939, \mn@doi [Physical Review] {10.1103/PhysRev.55.364}, \href {https://ui.adsabs.harvard.edu/abs/1939PhRv...55..364T} {55, 364}

\bibitem[\protect\citeauthoryear{{Tolos} \& {Fabbietti}}{{Tolos} \& {Fabbietti}}{2020}]{Tolos2020}
{Tolos} L.,  {Fabbietti} L.,  2020, \mn@doi [Progress in Particle and Nuclear Physics] {10.1016/j.ppnp.2020.103770}, \href {https://ui.adsabs.harvard.edu/abs/2020PrPNP.11203770T} {112, 103770}

\bibitem[\protect\citeauthoryear{Tolos, Centelles  \& Ramos}{Tolos et~al.}{2016}]{Tolos_2016}
Tolos L.,  Centelles M.,   Ramos A.,  2016, \mn@doi [\apj] {10.3847/1538-4357/834/1/3}, 834, 3

\bibitem[\protect\citeauthoryear{{Tolos}, {Centelles}  \& {Ramos}}{{Tolos} et~al.}{2017}]{Tolos_2017}
{Tolos} L.,  {Centelles} M.,   {Ramos} A.,  2017, \mn@doi [\pasa] {10.1017/pasa.2017.60}, \href {https://ui.adsabs.harvard.edu/abs/2017PASA...34...65T} {34, e065}

\bibitem[\protect\citeauthoryear{Traversi, Char  \& Pagliara}{Traversi et~al.}{2020}]{Traversi:2020aaa}
Traversi S.,  Char P.,   Pagliara G.,  2020, \mn@doi [\apj] {10.3847/1538-4357/ab99c1}, 897, 165

\bibitem[\protect\citeauthoryear{{Vinciguerra} et~al.,}{{Vinciguerra} et~al.}{2024}]{Vinciguerra24}
{Vinciguerra} S.,  et~al., 2024, \mn@doi [\apj] {10.3847/1538-4357/acfb83}, \href {https://ui.adsabs.harvard.edu/abs/2024ApJ...961...62V} {961, 62}

\bibitem[\protect\citeauthoryear{{Watts}}{{Watts}}{2019}]{Watts2019}
{Watts} A.~L.,  2019, in Xiamen-CUSTIPEN Workshop on the Equation of State of Dense Neutron-Rich Matter in the Era of Gravitational Wave Astronomy. p. 020008 (\mn@eprint {arXiv} {1904.07012}), \mn@doi{10.1063/1.5117798}

\bibitem[\protect\citeauthoryear{{Watts} et~al.,}{{Watts} et~al.}{2016}]{Watts_2016}
{Watts} A.~L.,  et~al., 2016, \mn@doi [Reviews of Modern Physics] {10.1103/RevModPhys.88.021001}, \href {https://ui.adsabs.harvard.edu/abs/2016RvMP...88b1001W} {88, 021001}

\bibitem[\protect\citeauthoryear{{Watts} et~al.,}{{Watts} et~al.}{2019}]{extp_watts}
{Watts} A.~L.,  et~al., 2019, \mn@doi [Science China Physics, Mechanics, and Astronomy] {10.1007/s11433-017-9188-4}, \href {https://ui.adsabs.harvard.edu/abs/2019SCPMA..6229503W} {62, 29503}

\bibitem[\protect\citeauthoryear{{Yang} \& {Piekarewicz}}{{Yang} \& {Piekarewicz}}{2020}]{YangPiek2020}
{Yang} J.,  {Piekarewicz} J.,  2020, \mn@doi [Annual Review of Nuclear and Particle Science] {10.1146/annurev-nucl-101918-023608}, \href {https://ui.adsabs.harvard.edu/abs/2020ARNPS..70...21Y} {70, 21}

\bibitem[\protect\citeauthoryear{Yue, Chen, Zhang  \& Zhou}{Yue et~al.}{2022}]{Yue:2021yfx}
Yue T.-G.,  Chen L.-W.,  Zhang Z.,   Zhou Y.,  2022, \mn@doi [Phys. Rev. Res.] {10.1103/PhysRevResearch.4.L022054}, 4, L022054

\bibitem[\protect\citeauthoryear{{Zhang} et~al.,}{{Zhang} et~al.}{2019}]{extp_zhang}
{Zhang} S.,  et~al., 2019, \mn@doi [Science China Physics, Mechanics, and Astronomy] {10.1007/s11433-018-9309-2}, \href {https://ui.adsabs.harvard.edu/abs/2019SCPMA..6229502Z} {62, 29502}

\makeatother
\end{thebibliography}
\bsp	
\label{lastpage}
\end{document}